\documentclass[12pt]{article}

\usepackage{jheppub}

\usepackage{amsmath}
\usepackage{mathtools}
\usepackage{amssymb}
\usepackage{amsfonts}
\usepackage{amsthm}
\usepackage{mathrsfs}
\usepackage[all]{xy}
\usepackage{tensor}
\usepackage[dvipsnames]{xcolor}
\usepackage[greek, english]{babel} 
\usepackage{teubner}  
\usepackage[normalem]{ulem} 
\usepackage{enumitem} 

\DeclareRobustCommand{\rchi}{{\mathpalette\irchi\relax}}
\newcommand{\irchi}[2]{\raisebox{\depth}{$#1\chi$}} 

\newcommand{\lie}{\pounds}
\newcommand{\Hi}{\mathcal{H}}
\newcommand{\PP}{\mathcal{P}}
\newcommand{\SSS}{\mathcal{S}}
\newcommand{\Ph}{\Phi}
\newcommand{\Th}{\Theta}
\newcommand{\Om}{\Omega}
\newcommand{\del}{\delta}
\newcommand{\al}{\alpha}
\newcommand{\dar}{\rchi}
\newcommand{\dy}{{\dar_Y}}
\newcommand{\dx}{{\dar}}
\newcommand{\hdx}{{\hat\dar}}
\newcommand{\hdy}{{\hat\dar_Y}}

\newcommand{\ep}{\epsilon}
\newcommand{\qo}{\text{\textqoppa}}
\newcommand{\ddiff}{{\text{Dir}_{\partial\Sigma} } } 

\newcommand{\beq}{\begin{equation}}
\newcommand{\eeq}{\end{equation}}

\begin{document}

\title{Local phase space and edge modes for diffeomorphism-invariant theories}
\author{Antony J. Speranza   }

\affiliation{Maryland Center for Fundamental Physics, University of Maryland, \\
\mbox{College Park,
MD 20742, USA}}
\emailAdd{asperanz@gmail.com}
\date{July 14, 2017}

\abstract{
We discuss an approach to characterizing local degrees of freedom of a subregion 
in diffeomorphism-invariant theories using the extended phase space of Donnelly and Freidel,
[JHEP 2016 (2016) 102].  
Such a characterization is important for defining local observables and entanglement
entropy in gravitational theories.  Traditional
phase space constructions for subregions are not invariant with respect
to diffeomorphisms that act at the boundary.  The extended phase space remedies this 
problem by introducing edge mode fields at the boundary whose transformations under
diffeomorphisms render the extended symplectic structure fully gauge invariant.  In this
work, we present a general construction for the edge mode symplectic
structure.  We show that the new fields satisfy a surface symmetry algebra 
generated by the Noether charges
associated with the edge mode fields.  For surface-preserving symmetries, the 
algebra is universal for all diffeomorphism-invariant theories, comprised of diffeomorphisms
of the boundary, $SL(2,\mathbb{R})$ transformations of the normal plane, and, in some cases,
normal shearing transformations.  We also show that if boundary conditions are chosen such
that surface translations are symmetries, the algebra acquires a central extension.   
}

\maketitle
\flushbottom

\section{Introduction}
In gravitational theories, the problem of defining local subregions and observables is complicated
by diffeomorphism invariance.  Because it is a gauge symmetry, diffeomorphism invariance
leads to constraints that must be satisfied by initial data for the field equations.  These constraints
relate the values of fields 
in one subregion of a Cauchy slice to their values elsewhere, 
so that the fields 
cannot be interpreted as observables localized to a particular region.   
While this is true in any gauge theory, a further challenge for diffeomorphism-invariant theories
is that specifying a particular subregion is nontrivial, since diffeomorphisms can change the 
subregion's coordinate position.  

A related issue in quantum gravitational theories is the problem of defining entanglement entropy
for a subregion.  The usual definition of entanglement entropy assumes a factorization of the 
Hilbert space $\Hi = \Hi_A\otimes \Hi_{\bar{A}}$ into tensor factors $\Hi_A$ and $\Hi_{\bar{A}}$
associated with a subregion $A$ and its complement $\bar{A}$. However, all physical states
in a gauge theory are required to be annihilated by the constraints, and the nonlocal relations
the constraints impose on the physical Hilbert space prevents such a factorization from occurring.
One way of handling this nonfactorization is to define the entropy in terms of the algebra of 
observables for the local subregion \cite{Casini2014a}.  This necessitates a choice of 
center for the algebra, which roughly corresponds to
 Wilson lines that are cut by the entangling surface.   
This procedure is further complicated in gravitational theories, since the local subregion 
and its algebra of observables must be defined in a diffeomorphism-invariant manner.  
Thus, the issues of local observables
and entanglement in gravitational theories are intertwined.  

Despite these challenges, there are indications that a well-defined notion of local observables
and entanglement should exist in gravitational theories.  Holography provides a compelling  example, 
where the entanglement of bulk regions bounded by an extremal surface may be expressed 
in terms of entanglement in the CFT via the Ryu-Takayanagi formula and its quantum corrections
\cite{Ryu:2006bv, Faulkner2013a}.
Such regions are defined relationally relative to a fixed region
on the boundary, and hence give a diffeomorphism-invariant characterization of the local subregion.
Work regarding bulk reconstruction suggests that the algebra of 
observables for this subregion is 
fully expressible in terms of the subregion algebra of the CFT 
\cite{Dong2016, Jafferis2015, Czech2012, Cotler2017, Harlow2016, Almheiri2015}. 

In addition, there are various pieces of circumstantial evidence suggesting that entanglement 
entropy is a well-defined and useful concept in quantum gravity.  
The gravitational field equations have been shown to follow from applying
the first law of entanglement
entropy \cite{Blanco2013a, Bhattacharya:2012mi}  
to subregions, both in holography \cite{Lashkari2013, Faulkner:2013ica, Swingle2014,
Faulkner2015, Faulkner2017} 
and for more general gravitational theories \cite{Jacobson2015a, Casini2016a, Speranza2016,
Bueno2017}, all of which is predicated on a well-defined notion for entanglement for the local 
subregion.  In fact, it is conjectured that connectivity of the spacetime manifold arises  from
entanglement between the microscopic degrees of freedom from which the gravitational theory
emerges \cite{VanRaamsdonk2010}.
Furthermore, entanglement entropy provides a natural explanation for the proportionality between
black hole entropy and horizon area 
\cite{Sorkin:2014kta, Bombelli1986, Srednicki1993a, Frolov1993}, 
while finessing the issue of entanglement
divergences through renormalization of the gravitational couplings 
\cite{Susskind1994, Jacobson1994a, Cooperman:2013iqr}.  
However, in the case of gauge theories, the matching between entanglement 
entropy divergences and the 
renormalization of gravitational couplings is subtle.  
The entropy computed using conical methods \cite{Callan1994a}
contains contact
terms \cite{Kabat1995, Fursaev1997, Solodukhin2015}, 
which are related to the presence of edge modes on the entangling surface.  
These arise as a consequence of the nonfactorization of the Hilbert space due to 
the gauge constraints.  
Only when  the 
entanglement from these edge modes 
is properly handled does the black hole entropy 
have a statistical interpretation in terms of a von Neumann entropy \cite{Donnelly2012,
Donnelly2014, Donnelly2015}.  

Recently, Donnelly and Freidel presented a continuum description of the edge modes that arise 
both in Yang-Mills  theory and general relativity \cite{Donnelly2016}.  Using covariant 
phase space techniques \cite{Witten1986, Crnkovic1987, Crnkovic:1987tz, Ashtekar1991}, 
they construct
a symplectic potential and symplectic form associated with a local subregion.  
These are expressed as local integrals of the fields and their variations over a Cauchy surface
$\Sigma$.  However, one finds that they are not fully gauge-invariant:
gauge transformations that are 
nonvanishing at the boundary $\partial\Sigma$ change the symplectic form by 
boundary terms.  Invariance is restored by introducing 
new fields  in a neighborhood
of the boundary, whose change under gauge transformations cancels the boundary term from the 
original symplectic form.  These new edge modes thus realize the idea that boundaries 
break gauge invariance, and cause some would-be gauge modes to become  
degrees of freedom associated with the subregion \cite{Carlip1994, Carlip1997}.

The analysis of diffeomorphism-invariant theories in \cite{Donnelly2016} was restricted to 
general relativity with vanishing cosmological constant.  However, the 
construction can be generalized to arbitrary diffeomorphism-invariant theories, and it is the 
purpose of the present work to show how this is done.  The  symplectic
potential for the edge modes can be expressed in terms of the Noether charge and the 
on-shell Lagrangian
of the theory, and the symplectic form derived from it has contributions from the edge modes 
only
at the boundary.  These edge modes come equipped with set of symmetry transformations,
and the symmetry algebra is represented on the phase space as a Poisson bracket algebra.  
The generators of the surface symmetries are given by the Noether charges associated with the 
transformations.  We find that for generic diffeomorphism-invariant theories, the transformations
that preserve the entangling surface generate the algebra 
$\text{Diff}(\partial\Sigma)\ltimes\left(SL(2,\mathbb{R})
\ltimes \mathbb{R}^{2\cdot(d-2)}\right)^{\partial\Sigma}$.  In certain cases, including general 
relativity, the algebra is reduced to $\text{Diff}(\partial\Sigma)\ltimes 
SL(2,\mathbb{R})^{\partial\Sigma}$, consistent with the results of \cite{Donnelly2016}.
Furthermore, for any other theory, there always exists a modification of the symplectic
structure in the form of a Noether charge ambiguity \cite{Jacobson1994b} that reduces the algebra
down to $\text{Diff}(\partial\Sigma)\ltimes 
SL(2,\mathbb{R})^{\partial\Sigma}$.  We also discuss what happens when the algebra
is enlarged to include surface translations, the transformations that do not map $\partial\Sigma$
to itself.  In order for these transformations to be Hamiltonian, the dynamical fields generically have 
to satisfy boundary conditions at $\partial\Sigma$.  Assuming the appropriate boundary
conditions can be found, the full surface symmetry algebra is a central extension of either
$\text{Diff}(\partial\Sigma)\ltimes\left( SL(2,\mathbb{R})\ltimes\mathbb{R}^2 \right)^{\partial\Sigma}$,
or a larger, simple Lie algebra.
The appearance of central charges in these algebras is familiar from similar constructions
involving edge modes at asymptotic infinity or black hole horizons 
\cite{Brown1986a, Carlip1997, Carlip2011}.

The construction of the extended phase space for arbitrary diffeomorphism-invariant theories
is useful for a number of reasons.  For one, higher curvature corrections
to the Einstein-Hilbert action generically appear  due to quantum gravitational effects.  
It is useful to have a formalism that can compute the corrections to the edge mode entanglement
coming from these higher curvature terms.  Additionally, there are several 
diffeomorphism-invariant theories that are simpler than general relativity in four dimensions, 
such as 2 dimensional dilaton gravity or 3 dimensional gravity in Anti-de-Sitter space.  
These could be useful testing grounds in which to understand the edge mode 
entanglement entropy, before trying to tackle the problem in four or higher dimensions.  
Finally, the general construction clarifies the relation of the extended phase space to the 
Wald formalism \cite{Wald1993a, Iyer1994a}, a connection that was also noted in
\cite{Geiller2017}.  

This paper begins with a review of the covariant phase space in section \ref{sec:cps}.
Care is taken to describe vectors and differential forms on this infinite-dimensional space, 
and also to understand the effect of diffeomorphisms of  the spacetime manifold on 
the covariant phase space.  Section \ref{sec:edge} discusses the $X$ fields that appear
in the extended phase space, which give rise to the edge modes.  Following this,
the construction of the extended phase space is given in section \ref{sec:eps}, which
describes how the edge mode fields contribute to the extended symplectic form.  Ambiguities
in the construction are characterized in section \ref{sec:jkm}, and the surface symmetry 
algebra is identified in section \ref{sec:ssa}.  Section \ref{sec:disc} gives a summary of results 
and ideas for 
future work.

\section{Covariant phase space} \label{sec:cps}
The covariant phase space \cite{Witten1986, Crnkovic1987, Crnkovic:1987tz, Ashtekar1991}
provides a Hamiltonian description of a field theory's degrees of freedom
while maintaining spacetime covariance.  This is achieved by working with the space $\SSS$ of 
solutions to the field equations.  As long as the field equations admit a well-posed initial 
value formulation, each solution is in one-to-one correspondence with its initial data
on some Cauchy slice.  $\SSS$ 
may therefore
be used to construct a phase space that is equivalent to other Hamiltonian formalisms, such
as ADM \cite{Arnowitt2008},
but  since it does not require a choice of Cauchy
slice and decomposition into spatial and time coordinates, spacetime covariance remains 
manifest.  The specification of a Cauchy surface and time variable can be viewed as a choice 
of coordinates on $\SSS$, with each solution being identified by its initial data.  

Working directly
with $\SSS$ allows   coordinate-free techniques to be applied  to both the 
spacetime manifold and the phase space itself.  In particular, the exterior calculus on the 
$\SSS$ gives a powerful language for describing the phase space symplectic geometry.  
We will follow the treatment of the exterior calculus given in 
\cite{Donnelly2016},\footnote{For an extended review of this formalism, see
\cite{Compere2007} and references therein.} where it was 
used to provide an extremely efficient way of identifying edge modes for a local subregion
in a gauge theory.   
This section provides a review of 
the formalism,
on which the remainder of this paper heavily relies.  

The theories under consideration consist of dynamical fields, including the metric $g_{ab}$ and 
any matter fields, propagating on a spacetime manifold $M$.  
These fields satisfy diffeomorphism-invariant equations of motion, and the phase space is 
constructed from the infinite-dimensional space of solutions to these equations, $\SSS$.  
Despite being infinite-dimensional, many concepts from finite-dimensional differential 
geometry, such as vector fields, one-forms, and Lie derivatives, 
extend straightforwardly to $\SSS$, assuming it satisfies some technical
requirements such as being a Banach manifold \cite{Lee1990a, Lang1985}.
One begins by understanding the functions on $\SSS$, 
a wide
class of which is provided by the dynamical fields themselves.  Given 
a spacetime point $x\in M$ and a field $\phi$, the function $\phi^x$ associates to each 
solution the value of $\phi(x)$  in that solution.  
More generally, functionals of the dynamical fields, such
as integrals over regions of spacetime, also define functions on $\SSS$ by simply evaluating 
the functional in a given solution.  
We will often denote $\phi^x$ simply by $\phi$, with the 
dependence on the spacetime point $x$ implicit. 

A vector at a point of $\SSS$ describes an infinitesimal displacement away from a particular
solution, and hence corresponds to a solution of the linearized field equations.  Specifying 
a linearized solution about each full solution then defines a vector field $V$ 
on all of $\SSS$.  
The vector field acts on $\SSS$-functions  as a directional derivative, and in particular its
action on the functions $\phi^x$ is to give a new function $\Ph_V^x\equiv V[\phi^x]$, 
which, given a solution, evaluates the linearization $\Ph$ of the field $\phi$ 
at the point $x$.  This also allows us to define the exterior derivative of the 
functions $\phi^x$, denoted $\delta \phi^x$.  When contracted with the vector field
$V$, the one-form $\delta\phi^x$ simply returns the scalar function $\Ph_V^x$.  The 
one-forms $\del\phi^x$ form an overcomplete basis, so that arbitrary one-forms may be 
expressed as sums (or integrals over the spacetime point $x$) of $\del\phi^x$.  
This basis is overcomplete because the functions $\phi^x$ at different
points $x$ are related through the equations of motion, so that the forms $\del\phi^x$ are 
related as well.

Forms of higher degree can be constructed from the $\del\phi^x$ one-forms by 
taking exterior products.  The exterior product of a $p$-form $\alpha$ and a $q$-form $\beta$
is simply written $\alpha\beta$, and satisfies $\alpha\beta = (-1)^{pq} \beta\alpha$.  Since we only
ever deal with exterior products of forms defined on $\SSS$ instead of more general tensor
products, no ambiguity arises by omitting the $\wedge$ symbol, which we instead reserve 
for spacetime 
exterior products.  The action of the exterior derivative on arbitrary forms is fixed as usual by 
its action on scalar functions, along with the requirements of linearity, nilpotency $\delta^2=0$, 
and that it acts as an antiderivation,
\beq
\delta(\alpha\beta) = (\delta\alpha)\beta + (-1)^p \alpha \delta\beta.
\eeq
The exterior derivative $\delta$ always increases the degree of the form by one.  On the other 
hand, each vector field
$V$ defines an antiderivation $I_V$ that reduces the degree by one through contraction.  
$I_V$ can be completely characterized by its action on one-forms $I_V \del\phi^x=\Phi_V^x$,
along with the antiderivation property, linearity, nilpotency $I_\Ph^2
=0$, and requiring that it annihilate scalars.  
Just as in finite dimensions, the action of the $\SSS$ Lie derivative, denoted $L_V$, 
is related to $\del$ and $I_V$ via Cartan's magic formula \cite{Lang1985}
\beq \label{eqn:cartanmagic}
L_V = I_V \delta + \delta I_V.
\eeq
$L_V$ is a derivation, $L_V (\alpha\beta) = (L_V\alpha)\beta + \alpha L_V \beta$, that 
preserves the degree of the form.

We next discuss the consequences of working with  diffeomorphism invariant theories.  
A diffeomorphism $Y$ is a smooth, invertible map, $Y:M\rightarrow M$, sending the 
spacetime manifold $M$ to itself.  The diffeomorphism  induces a map of
tensors at $Y(x)$ to tensors at $x$ through the pullback $Y^*$ \cite{Wald1984}. 
Diffeomorphism invariance is simply the statement that if a configuration of tensor fields
$\phi$ satisfy the equations of motion, then so do the pulled back fields $Y^*\phi$. 
Now consider a one-parameter family of diffeomorphisms $Y_\lambda$, with $Y_0$ the identity.  
This yields a family of fields $Y_\lambda^*\phi$ that all satisfy the equations of motion.  The first
order change induced by $Y^*_\lambda$ defines the spacetime Lie derivative $\lie_\xi$ 
with respect to $\xi^a$, the tangent vector to the flow of $Y_\lambda$.
Consequently, $\lie_\xi \phi$ must be a solution to the linearized field 
equations, and the infinitesimal diffeomorphism generated by $\xi^a$ defines a vector field on
$\SSS$, which we  denote $\hat{\xi}$, 
whose action on $\delta\phi$ is 
\beq
I_{\hat\xi} \delta \phi\equiv \lie_\xi \phi.
\eeq

The diffeomorphisms we have considered so far have been taken to act the same on all solutions.
A useful generalization of this are the solution-dependent diffeomorphisms, defined through a 
function,
$\mathscr{Y}:\SSS\rightarrow \text{Diff}(M)$, valued in the diffeomorphism
group of the manifold, $\text{Diff}(M)$.  
Letting $Y$ denote the image of this function, 
we would like
to understand how the Lie derivative $L_V$ and exterior derivative $\del$ on $\SSS$ 
combine with the action of the pullback $Y^*$.  
In the
case $\mathscr{Y}$ is constant on $\SSS$, the Lie derivative simply commutes with $Y^*$, and so
$L_V Y^*\al = Y^* L_V \al$, where $\al$ is any form constructed from fields
and their variations at a single spacetime point.  When $Y$ is not constant, 
$V$ generates one-parameter families of diffeomorphisms $Y_\lambda$ and 
forms $\alpha_\lambda$ along the flow in $\SSS$.  At a given solution $s_0$, define a 
solution-independent diffeomorphism  
$Y_0 \equiv \mathscr{Y}(s_0)$ by the value of $\mathscr{Y}$ at $s_0$.
Then $Y^*_\lambda \alpha_\lambda$ and $Y^*_0\alpha_\lambda$ are  related
to each other at all values of $\lambda$ by a diffeomorphism, $Y_\lambda^* (Y_0^{-1})^*$.  
The first order change in these quantities at $\lambda=0$ is given by 
$L_V$, and since the two quantities differ at first order by an infinitesimal diffeomorphism, we find 
\beq \label{eqn:LphY*a}
L_V Y^*\al = L_V Y_0^*\al + Y^* \lie_{\dar(Y;V)} \al = Y^*(L_V \al + \lie_{\dar(Y;V)} \al).
\eeq
It is argued in appendix \ref{app:ids}, identity \ref{id:LVY*a},
that the vector $\dar^a(Y;V)$ depends linearly on 
$V$, and hence defines a one-form on $\SSS$, denoted $\dar_Y^a$.\footnote{In
\cite{Donnelly2016}, $\dar_Y^a$ was denoted $\del_Y^a$.  We choose a different notation
to emphasize
that $\dar_Y^a$ is not an exact form, and to avoid confusion with the exterior derivative $\del$.}
This yields the pullback formula for $L_V$,
\beq
L_V Y^*\al = Y^*(L_V\al + \lie_{I_V \dy} \alpha).
\eeq
Applying (\ref{eqn:cartanmagic}) to this equation, one can derive the pullback
formula for exterior derivatives  from \cite{Donnelly2016} (see \ref{id:dY*a} for details),
\beq \label{eqn:dY*a}
\delta Y^*\alpha = Y^*(\delta\al+\lie_{\dy}\al).
\eeq

A number of properties of the variational vector field $\dar_Y^a$ follow from the 
formulas above.  First, note $\dar_Y^a$ is not an exact form on $\SSS$; rather, 
its exterior derivative
can be deduced from (\ref{eqn:dY*a}),
\beq
0 = \delta\del Y^*\al = Y^*(\del \lie_\dy\al +\lie_\dy \del\al +\lie_\dy \lie_\dy \al) = Y^*(\lie_{\del
(\dy)} \al + \lie_\dy \lie_\dy\al),
\eeq
and applying \ref{id:liedarliedar}, we conclude
\beq \label{eqn:ddx}
\delta(\dy)^a = -\frac12[\dy,\dy]^a.
\eeq
Another useful formula relates $\dar_Y^a$ to the vector $\dar_{Y^{-1}}^a$ associated with the 
inverse of $Y$.  Using that $Y^*$ and $(Y^{-1})^*$ are inverses of each other, we find
\beq
\delta \al = \delta Y^* (Y^{-1})^*\al = Y^*[\delta(Y^{-1})^*\al + \lie_{\dy} (Y^{-1})^* \al]
= \del\al +\lie_{\del_{Y^{-1}}}\al + \lie_{Y^*\dy} \al,
\eeq
where the last equality involves the identity \ref{id:liexiYi}. This implies
\beq \label{eqn:dy-1}
\dar_{Y^{-1}}^a = -Y^*\dar_Y^a.
\eeq
Additional identities are derived in appendix \ref{app:ids}.

Finally, as a spacetime vector field, $\dar_Y^a$ also defines a vector-valued one-form 
$\hdy$ on $\SSS$, which acts as $I_{\hdy} \delta\phi = \lie_\dy \phi$.  The contraction
$I_{\hdy}$ defines a derivation that preserves the degree of the form, 
in contrast to $I_{\hat\xi}$, 
which is an antiderivation that reduces the degree.  Similarly, $\delta(\dy)^a$
defines a vector-valued two-form on $\SSS$, and produces an antiderivation $I_{\delta(\dy)\,\hat{}}$
that increments the degree.

\section{Edge mode fields} \label{sec:edge}
Edge modes appear when a gauge symmetry is broken due to the presence of a 
boundary $\partial\Sigma$ of a Cauchy surface $\Sigma$.  
The classical phase space  or quantum mechanical Hilbert space associated with 
$\Sigma$ transforms nontrivially under gauge transformations that act at the boundary.  
This can be understood from the perspective of Wilson loops that are cut by the 
boundary.   A closed Wilson loop is gauge-invariant, but the cut Wilson loop
becomes a Wilson line in $\Sigma$, whose endpoints transform in some representation
of the gauge group.  To account for these cut-Wilson-loop degrees of freedom, one can 
introduce fictitious charged fields at $\partial \Sigma$, which can be attached to the 
ends of the Wilson lines to produce a gauge-invariant object.   These new fields are the 
edge modes of the local subregion.  They account for the possibility of charge 
density existing outside of $\Sigma$, which would affect the fields in $\Sigma$ due to 
Gauss law constraints.  The contribution of the edge modes to the entanglement can 
therefore be interpreted as parameterizing  ignorance of such localized charge densities
away from $\Sigma$.

A similar picture arises in the classical phase space of a diffeomorphism-invariant theory.  
The edge modes appear when attempting to construct a symplectic
structure  associated with $\Sigma$ for the solution space $\SSS$.  
Starting with the Lagrangian of the theory, one can construct from its variations a 
symplectic current $\omega$, a spacetime $(d-1)$-form whose integral over a spatial 
subregion $\Sigma$ provides a candidate presymplectic form. 
However, this form fails to be 
diffeomorphism invariant for two reasons.  First, a diffeomorphism moves points on the mainfold
around, and hence changes the shape and coordinate location of the 
surface.  
Second, since solutions related to each other by a diffeomorphism  represent
the same physical 
configuration,  the true phase space $\PP$ is obtained by projecting all solutions in 
a  gauge orbit in $\SSS$ down to a single representative.  In order for the symplectic form
to be compatible with this projection, the infinitesimal diffeomorphisms must be  degenerate 
directions of the presymplectic form \cite{Lee1990a}.  
This is equivalent to saying that the Hamiltonian generating the diffeomorphism may be chosen
to vanish. 
While the symplectic form obtained by integrating $\omega$ over a surface 
is degenerate for diffeomorphisms that vanish sufficiently quickly at its boundary, those that do not
produce boundary terms that spoil degeneracy.

As demonstrated in \cite{Donnelly2016}, these problems can be handled by introducing a collection
of additional fields $X$ whose contribution to the symplectic form restores diffeomorphism 
invariance.  These fields are the edge modes of the extended phase space.  This section
 is devoted to describing these fields and their 
 transformation properties under diffeomorphisms; 
the precise way in which they contribute to the symplectic form is discussed in section
\ref{sec:eps}.  

The fields $X$ can be defined through a $\text{Diff($M$)}$-valued function $\mathscr{X}:\SSS
\rightarrow \text{Diff}(M)$.  In a given solution $s$, $X$ is identified with 
the diffeomorphism in the image of the map, 
$X=\mathscr{X}(s)$.  One way to interpret $X$ is as defining
a map from (an open subset of) $\mathbb{R}^d$ into the spacetime manifold $M$,
and hence can be thought of as a choice of coordinate system on covering the local
subregion $\Sigma$.\footnote{We assume
for simplicity that the subregion of interest can be covered by a single coordinate
system.  For topologically nontrivial subregions, the fields may consist of a collection of maps
$X_i$, one for each coordinate patch needed to cover the region.}  
A full solution
to the field equations now consists of specifying the map $X$ as well as the value of the dynamical
fields $\phi(x)$ at each point in spacetime.  
The transformation law for $X$ 
under a diffeomorphism $Y:M\rightarrow M$ is given by the pullback along
$Y^{-1}$,  $\bar{X} = Y^{-1} \circ X$.   

Since $X$ defines a diffeomorphism from $\mathbb{R}^d$ to $M$, it can be used to pull back 
tensor fields  on $M$ to $\mathbb{R}^d$.  We can argue as before that the Lie derivative $L_V$
and exterior derivative $\delta$ satisfy pullback forumlas analogous to equations (\ref{eqn:LphY*a})
and (\ref{eqn:dY*a}),
\begin{align}
L_V X^*\al &= X^*(L_V \al+\lie_{I_V \dar_X} \al) \label{eqn:LphX*a} \\
\delta X^*\al &= X^*(\del \al + \lie_{\dar_X} \al),  \label{eqn:dX*a}
\end{align}
which serve as defining relations for the variational spacetime vector $\dar_X^a$.  
The result of contracting $\dar_X^a$
with a vector field $\hat\xi$ corresponding to a spacetime 
diffeomorphism
 can be deduced by first noting that the pulled back fields $X^*\phi$ are invariant under
 diffeomorphisms, since 
\beq \label{eqn:barX*}
\bar{X}^* Y^*\phi = X^*(Y^{-1})^* Y^*\phi = X^*\phi.  
\eeq
In particular,
the $\SSS$ Lie derivative $L_{\hat\xi}$ must annihilate $X^* \phi$ for any $\xi$, so from 
(\ref{eqn:LphX*a}),
\beq \label{eqn:LxiX*phi}
0=L_{\hat\xi} X^*\phi = X^*(L_{\hat\xi} \phi +\lie_{I_{\hat{\xi}}\dar_X}\phi) = X^*(\lie_\xi \phi+ 
\lie_{I_{\hat\xi} \dar_X} \phi),
\eeq
and hence
\beq \label{eqn:Ixidar}
I_{\hat\xi} \dar_X^a = -\xi^a.
\eeq

We can also derive the transformation law for $\dar_X^a$ under a diffeomorphism from the 
pullback formulas (\ref{eqn:dY*a}) and (\ref{eqn:dX*a}).  On the one hand we have 
\beq
\delta \bar{X}^*\al = \bar{X}^*(\del \al + \lie_{\dar_{\bar{X}}} \al),
\eeq
while on the other hand this can also be computed as 
\beq
\del\bar{X}^*\al = \del X^* (Y^{-1})^* \al = X^*[\del (Y^{-1})^* \al + \lie_{\dar_X} (Y^{-1})^* \al]
= \bar{X}^*(\del \al  + \lie_{\dar_{Y^{-1}}} \al + \lie_{Y^*\dar_X} \al)
\eeq
where the last equality employed identity \ref{id:liexiYi}.  Comparing these expressions
and applying the formula (\ref{eqn:dy-1}) for $\dar_{Y^{-1}}^a$ gives the transformation law
\beq\label{eqn:dbarXa}
\dar_{\bar{X}}^a = Y^*(\dar_X^a-\dar_Y^a).
\eeq

The $X$ fields lead to an easy prescription for forming diffeomorphism-invariant quantities: simply
work with the pulled back fields $X^*\phi$.  These are diffeomorphism-invariant due to equation
(\ref{eqn:barX*}), and consequently the variation $\delta X^*\phi$ is as well.  We can explicitly 
confirm that $\del X^*\phi$ are annihilated by infinitesimal diffeomorphisms $\hat\xi$:
\beq
I_{\hat\xi}\del X^*\phi = I_{\hat\xi}X^*(\del\phi + \lie_{\dar_X}\phi) = X^*(\lie_\xi\phi - \lie_\xi\phi)=0.
\eeq
Another combination of one-forms that appears frequently is $\alpha + I_{\hat\dar_X}\alpha$,
and it is easily checked that $I_{\hat\xi}$ annihilates this sum.  Finally, we note that when
no confusion will arise, we will simply denote $\dar_X^a$ by $\dar^a$ to avoid excessive
clutter.  When referring to other diffeomorphisms besides $X$, we will explicitly include the 
subscript, as in $\dar_Y^a$.

\section{Extended phase space} \label{sec:eps}
We now turn to the problem of defining a gauge-invariant symplectic form to associate with 
the local subregion $\Sigma$.  
The standard procedure of \cite{Lee1990a, Wald1993a, Iyer1994a}
begins with a Lagrangian $L[\phi]$, 
a spacetime $d$-form constructed covariantly from the dynamical fields $\phi$.  Its variation 
takes the form
\beq \label{eqn:dL}
\delta L = E\cdot\delta\phi + d\theta,
\eeq
where $E=0$ are the dynamical field equations, and the exact form $d\theta$, where 
$d$ denotes the spacetime exterior derivative, defines the 
symplectic potential current $(d-1)$-form $\theta\equiv\theta[\phi; \delta\phi]$, which is a one-form
on solution space $\mathcal{S}$.  The $\SSS$-exterior derivative of $\theta$ defines the symplectic
current $(d-1)$-form, $\omega=\delta\theta$, whose integral over $\Sigma$ normally
defines the presymplectic form $\Om_0$ for the phase space.  As a consequence of 
diffeomorphism-invariance, $\Om_0$ contains degenerate directions: it annihilates 
any infinitesimal diffeomorphism generated by vector field $\xi^a$
that vanishes sufficiently quickly near the boundary.  This is succinctly expressed 
for such a vector field by 
$I_{\hat\xi}\Om_0=0$.  The true phase space $\PP$  
is obtained by quotienting out these degenerate directions by mapping all 
diffeomorphism-equivalent solutions to a single point in $\PP$.  $\Omega_0$ then defines a 
nondegenerate symplectic form on $\PP$ through the process of phase space reduction
\cite{Lee1990a}.  

This procedure is deficient for a local subregion $\Sigma$ because $\Om_0$
fails to be degenerate for diffeomorphisms that act near the boundary $\partial\Sigma$.  
If the boundary were at asymptotic infinity, such diffeomorphisms could be disallowed by
imposing boundary conditions on the fields, 
or could otherwise be regarded as true
time evolution with respect to the fixed asymptotic structure, 
in which case degeneracy would 
not be expected \cite{Ashtekar1991}.  For a local subregion, however, neither option is acceptable.   
Imposing a boundary condition on the fields at $\partial\Sigma$ has a nontrivial effect on the 
dynamics \cite{Andrade2015b, Andrade2015a, Andrade2016}, whereas we are interested in a  phase space that locally reproduces the same dynamics
as the theory defined on the full spacetime manifold $M$.  
Furthermore, the diffeomorphisms acting
at $\partial\Sigma$ cannot be regarded at true time evolution generated by a nonvanishing 
Hamiltonian, because these diffeomorphisms  are degenerate directions of a presymplectic 
form for the entire manifold $M$.  

Donnelly and Freidel \cite{Donnelly2016}
proposed a resolution to this issue by extending
the local phase space to 
include the $X$
fields described in section  \ref{sec:edge}.  The minimal prescription for introducing
them into the theory is to simply replace the Lagrangian with its pullback $X^*L$.  Since the 
Lagrangian is a covariant functional of the fields, $X^*L[\phi]=L[X^*\phi]$, so that the 
pulled back Lagrangian
depends only on the redefined fields $X^*\phi$, and is otherwise independent of $X$.  The
variation of this Lagrangian gives
\beq \label{eqn:dLX*phi}
\delta L[X^*\phi] = E[X^*\phi] \cdot \delta X^*\phi + d\theta[X^*\phi; \delta X^*\phi].
\eeq
Thus the redefined fields satisfy the same equations of motion $E[X^*\phi]=0$ as the original
fields, and, due to diffeomorphism invariance, this implies that the original $\phi$ fields must 
satisfy the equations as well.  Additionally,  the Lagrangian had no further dependence
on $X$, which means the $X$ fields do not satisfy any field equations.  
 If $X$ is understood as defining a coordinate system
for the local subregion, the dynamics of the extended $(\phi,X)$ system is simply given by the 
original field equations, expressed in an arbitrary coordinate system determined by $X$.

The symplectic potential current is read off from (\ref{eqn:dLX*phi}),
\beq\label{eqn:th'}
\theta' = \theta[X^*\phi; \delta X^*\phi] = \theta[X^*\phi; X^*(\delta\phi + \lie_{\dx}\phi)]
= X^*(\theta+I_{\hdx}\theta).
\eeq
This object is manifestly invariant with respect to solution-dependent diffeomorphisms, since 
both $X^*\phi$ and $\delta X^*\phi$ are.  In particular, $\theta'$ annihilates any 
infinitesimal diffeomorphism $I_{\hat\xi}$, as a consequence of the fact that $I_{\hat\xi}
\del X^*\phi = 0$ (see equation \ref{eqn:LxiX*phi}).  An equivalent 
expression for $\theta'$ can be obtained by introducing the Noether current 
for a vector field $\xi^a$,
\beq
J_\xi = I_{\hat\xi} \theta-i_\xi L,
\eeq
where $i_\xi$ denotes contraction with the spacetime vector $\xi^a$.
Due to diffeomorphism invariance, 
$J_\xi$ is an exact form when the equations of motion hold \cite{Wald1993a, Iyer1994a}, 
and may be 
written
\beq
J_\xi = dQ_\xi + C_\xi,
\eeq 
where $Q_\xi$ is the Noether charge and 
$C_\xi=0$ are combinations of the field equations that comprise the constraints for the 
theory \cite{Iyer1995b}.  Then $\theta'$ in (\ref{eqn:th'}) may be expressed
on-shell
\beq\label{eqn:th'2}
\theta' = X^*(\theta+i_{\dx} L + d Q_{\dx}).
\eeq

As an aside, note that we can vary the Lagrangian with respect to $(\phi, X)$ instead of 
the redefined fields $(X^*\phi, X)$, and equivalent dynamics arise.  This variation
produces  
\beq \label{eqn:dX*L}
\delta X^*L[\phi] = X^*(\delta L+\lie_{\dx}L) = X^*(E\cdot\del\phi) +d X^*(\theta+i_{\dx}L),
\eeq
where Cartan's magic formula $\lie_\dx = i_\dx d + d i_\dx$ was used, along with the fact that $d$
commutes with pullbacks.
Again, $\phi$ satisfies the same field equation $E[\phi]=0$, and $X$ is subjected to no
dynamical equations.  This variation suggests a  potential current
$\theta''=X^*(\theta
+i_{\dx}L)$, which differs from (\ref{eqn:th'2}) by the exact form $dX^*Q_\dx$.  This 
difference is simply an ambiguity in the definition of the  potential current, since shifting 
it by an exact form does not affect equation (\ref{eqn:dL}) \cite{Jacobson1994b, Iyer1994a}.
  However,
$\theta''$ does not annihilate infinitesimal diffeomorphisms $I_{\hat\xi}$, making $\theta'$ the 
preferred choice.  The degeneracy requirement for the symplectic potential current therefore
gives a prescription to partially fix its ambiguities \cite{Geiller2017}, although additional
ambiguities remain, and are discussed in section \ref{sec:jkm}.

The symplectic potential $\Theta$ 
is now constructed by integrating $\theta'$ over the local subregion.  
Since $\theta'$ is defined as a pullback by $X^*$, its integral must be over the pre-image $\sigma$, 
for which  $X(\sigma)=\Sigma$.  This gives
\begin{align}
\Theta &= \int_\sigma \theta[X^*\phi;\delta X^*\phi] \label{eqn:Ths} \\
&= \int_\Sigma(\theta+i_{\dx}L) + \int_{\partial\Sigma} Q_{\dx}\label{eqn:ThS}.
\end{align}
The second line uses the alternative expression (\ref{eqn:th'2}) for 
$\theta'$, and is written as an integral 
of fields defined on the
original local subregion $\Sigma$, without pulling back by $X$.   This makes use of the general
formula 
$\int_\sigma X^*\alpha = \int_{X(\sigma)} \alpha$, and also applies  
Stoke's theorem $\int_\Sigma d\alpha
= \int_{\partial\Sigma}\alpha$ to write the Noether charge as a boundary integral.
Equation (\ref{eqn:ThS}) differs from the symplectic potential for the nonextended phase space,
$\Theta_0 = \int_\Sigma\theta$, by both a boundary term depending on the Noether charge, as 
well as a bulk term coming from the on-shell value of the Lagrangian.  For vacuum
general relativity with no cosmological constant, this extra bulk contribution vanishes, being
proportional to the Ricci scalar \cite{Donnelly2016}.  
However, when matter is present or the cosmological constant
is nonzero, this extra bulk contribution to $\Theta$ can survive.  As we  discuss below,
this bulk term imbues the symplectic form on the reduced phase space $\PP$ 
with nontrivial cohomology. 

Taking an exterior derivative of $\Th$ yields the symplectic form, $\Omega=\delta\Theta$.  The expression
(\ref{eqn:Ths}) leads straightforwardly to
\beq\label{eqn:Oms}
\Om = \int_\sigma \omega[X^*\phi; \delta X^*\phi, \delta X^*\phi], 
\eeq
where we recall the definition of the symplectic current $\omega=\delta\theta$.  This expression
for $\Omega$ makes it clear that it is invariant with respect to all diffeomorphisms, and  
that infinitesimal diffeomorphisms are degenerate directions, again because $I_{\hat\xi}\delta
X^*\phi = 0$.  
The symplectic form can also be expressed as an integral over $\Sigma$ and its boundary
using the original fields $\phi$, by computing the exterior derivative of (\ref{eqn:ThS}).   
Noting that the integrands implicitly involve a pullback by $X^*$, we find
\beq
\Om = \int_\Sigma(\omega +\lie_\dx \theta+\del i_\dx L+\lie_\dx i_\dx L) 
+\int_{\partial\Sigma} (\delta Q_{\dx} + \lie_\dx Q_\dx)
\eeq
The first term is the symplectic form for 
the nonextended theory, $\Omega_0 = \int_\Sigma \omega$.
The remaining three terms in the bulk $\Sigma$ integral simplify to an exact form on-shell 
 $d(i_\dx \theta +\frac12 i_\dx i_\dx L)$ (see identity \ref{id:liedxth}),  so the final expression
 is
 \beq\label{eqn:OmS}
 \Om = \int_\Sigma\omega + \int_{\partial\Sigma} \left[\delta Q_{\dx}+\lie_\dx Q_\dx + i_\dx\theta
 +\frac12i_\dx i_\dx L\right].
 \eeq 
 
Hence, we arrive at the important result that the symplectic form differs from $\Omega_0$
by terms localized on the boundary $\partial\Sigma$ involving $\dx^a$.  This immediately
implies that $\Omega$ has degenerate directions: any phase space vector field $V$ that 
vanishes on $\delta \phi$ and whose contraction with $\dx^a$ vanishes sufficiently
quickly near $\partial \Sigma$ will annihilate $\Omega$.  In fact, only the values of $\dx^a$
and  $\nabla_b\dx^a$ 
 at $\partial\Sigma$ contribute to (\ref{eqn:OmS}); all other freedom in $\dx^a$ is pure gauge.
To see why these are the only relevant pieces of $\dx^a$ for the symplectic form, 
we can use
the 
explicit expression for the Noether charge given in \cite{Iyer1994a}.  Up to ambiguities which
are discussed in section \ref{sec:jkm}, the Noether charge is given by
\beq\label{eqn:Qxi}
Q_\xi = -\ep_{ab}E\indices{^a^b^c_d}\nabla_{c}\xi^{d} + W_c\xi^c,
\eeq
where $\ep_{ab}$ is the spacetime volume form with all but the first two indices suppressed,
$E^{abcd}=\frac{\delta\mathcal{L}}{\delta R_{abcd}}$ is the variational derivative of the Lagrangian
scalar $\mathcal{L}=-(*L)$ 
with respect to the Riemann tensor, and inherits the index symmetries of the Riemann
tensor, and $W_c[\phi]$ is a tensor with $(d-2)$ covariant, antisymmetric indices suppressed, 
constructed locally from the dynamical fields; its precise form
is not needed in this work.

The last two terms in (\ref{eqn:OmS}) depend only on the value of $\dx^a$ on $\partial \Sigma$,
while the terms involving $Q_\dx$ can depend on 
derivatives of $\dx^a$.  From  (\ref{eqn:Qxi}), $Q_\dx$ 
involves one  derivative of $\dx^a$, and 
(\ref{eqn:OmS}) has terms involving the derivative of $Q_\dx$, so  that up to two 
derivatives of $\dx^a$ could contribute to the symplectic form.  
To see how these derivatives appear, we decompose $\delta Q_\dx$ as
\beq \label{eqn:dQdX}
\del Q_\dx =  Q_{\del (\dx)} + \qo_\dx,
\eeq
where $\qo_\xi=\qo_\xi[\phi;\del\phi]$\footnote{$\qo$ is the archaic Greek letter ``qoppa.''} 
is a variational one-form depending on a vector $\xi$ (which can be a differential form
on $\SSS$), given by
\beq\label{eqn:qoxi}
\qo_\xi = -\del(\ep_{ab} E\indices{^a^b^c_d})\nabla_c \xi^d-\ep_{ab}E\indices{^a^b^c_d} 
\del\Gamma^d_{ce}
\xi^e + \del W_c \xi^c,
\eeq
and $\del \Gamma^d_{ce}$ is the variation of the Christoffel symbol,
\beq
\del\Gamma^d_{ce} = \frac12g^{df}(\nabla_c\del g_{fe}+\nabla_e\del g_{fc}-\nabla_f \del g_{ce}).
\eeq
This decomposition is useful because $\qo_\dx$ contains only first derivatives of $\dx^a$, while
$Q_{\del\dx} = -\frac12 Q_{[\dx,\dx]}$ involves second derivatives through the derivative of the 
vector field Lie bracket.  

In appendix \ref{app:edge}, 
it is argued that 
the second derivatives of $\dx^a$ in $Q_{\del(\dx)}+\lie_\dx Q_\dx$ cancel out, 
 so
that the boundary contribution in (\ref{eqn:OmS}) 
depends on only $\dx^a$ and $\nabla_b \dx^a$ at $\partial\Sigma$. 
This means that $\Om$ has a large number of degenerate directions, corresponding to all
values of $\dx^a$ on $\Sigma$ that are not fixed by the values of $\dx^a$ and $\nabla_b\dx^a$
at the boundary.  The true phase space $\PP$ is then obtained by quotienting out these 
pure gauge degrees of freedom.  
In doing so, $\Omega$ descends 
to a nondegenerate, closed two-form on the quotient space \cite{Lee1990a}.
However, the symplectic potential 
$\Theta$ does not survive this projection.   It depends nontrivially on the value of $\dx^a$ 
everywhere on $\Sigma$ through the term involving the Lagrangian in (\ref{eqn:ThS}), 
which causes
it to become a multivalued form on the quotient space.  
One way to see its multivaluedness is to note that $i_\dx L$ is a top rank 
form on $\Sigma$, so, by the Poincar\'e
lemma applied to $\Sigma$, it can be expressed as the exterior derivative of a $(d-2)$-form,
\beq \label{eqn:dhX}
i_\dx L\big|_\Sigma = d h_X i_\dx L.
\eeq 
Here, $h_X$ is the homotopy operator that inverts the exterior derivative $d$ on closed forms on
$\Sigma$ \cite{Edelen2005}.  
As the notation suggests, it depends explicitly on the value of the $X$ fields throughout
$\Sigma$, which we recall can be thought of as defining a coordinate system for the subregion.
Since $h_X i_\dx L$ is a 
 spacetime $(d-2)$-form 
 and an $\SSS$ one-form, evaluated 
at $\partial\Sigma$ it may be expressed in terms of $\dx^a$ and $\del\phi$ at $\partial\Sigma$,
which provide a basis for local variational forms.  Hence,
\beq \label{eqn:inthX}
\int_\Sigma i_\dx L = \int_{\partial\Sigma} h_X i_\dx L,
\eeq
and we see that this latter expression depends on $\dx^a$ at $\partial\Sigma$, so therefore
will project to the quotient space.  However, $h_X$ will be a different operator depending on the 
values of the  
$X$ fields on $\Sigma$, and hence this boundary integral will give a different form on the 
reduced phase space for different bulk values of $X$.  This shows that the Lagrangian 
term in $\Theta$
projects  to a multivalued form on the quotient space. 

The failure of $\Theta$ to be single-valued implies that the reduced phase space $\PP$ 
has nontrivial
cohomology.  In particular, the projected 
symplectic form $\Omega$ is not exact, despite being closed.  For a given choice of the value of $\Theta$, the equation 
$\Om = \del\Theta$ still holds locally near a given solution 
in the reduced phase space, but there can be global 
obstructions since $\Theta$ may not return to the same value after tracing out a closed loop
in the solution space.  It would be interesting to investigate the consequences of this nontrivial
topology of the reduced phase space, and in particular whether it has any relation to the 
appearance of central charges in the surface symmetry algebra.  

Finally, note that for vacuum general relativity with no cosmological constant, 
the Lagrangian vanishes on shell, being proportional to the Ricci scalar.  In this special 
case, $\Theta$ is not multivalued and descends to a well-defined one-form on the 
reduced phase space, suggesting that the phase space topology simplifies. However, the 
inclusion of a cosmological constant or the presence of matter anywhere in the local subregion
leads back to the generic case in which $\Theta$ is multivalued.

\section{JKM ambiguities} \label{sec:jkm}
The constructions of the symplectic potential current $\theta$ and Noether charge $Q_\xi$ 
are subject to a number of ambiguities identified by Jacobson, Kang and Myers (JKM) 
\cite{Jacobson1994b, Iyer1994a}.  
These ambiguities correspond to the ability to add an exact form to 
the Lagrangian $L$, the potential current $\theta$, or the Noether charge $Q_\xi$ without 
affecting the dynamics or the defining properties of these forms.  Normally it is required that 
the ambiguous terms be locally constructed from the dynamical fields in a spacetime-covariant
manner.  In the extended phase space, however, there is additional freedom provided by the $X$
fields as well as the surfaces $\Sigma$ and $\partial\Sigma$ to construct forms that would 
otherwise fail to be  covariant.  The freedom provided by the $X$ fields is 
considerable, given that they can be used to construct homotopy operators as in (\ref{eqn:dhX})
and (\ref{eqn:inthX}) that mix the local dynamical fields $\phi$ at different spacetime points.  
For this reason, we refrain from using the $X$ fields in such an explicit manner to construct
ambiguity terms.  However, we allow for ambiguity terms that are constructed using the 
structures provided by $\Sigma$ and $\partial \Sigma$, such as their induced metrics and extrinsic
curvatures.  This allows for a wider class of Noether charges, including those that appear in 
holographic entropy functionals and the second law of black hole mechanics 
for higher curvature  theories \cite{Dong2014, Camps2013,
Miao2014, Wall2015}.

A simple example of which types of objects are permitted in constructing the ambiguity terms is 
provided by the unit normal $u_a$  to $\Sigma$ versus the lapse function $N$.  
Interpreting $X^\mu$ as a coordinate system for the local subregion, we can take $\Sigma$ to 
lie at $X^0=0$.  Then the lapse and unit normal are related by
\beq
u_a = - N \nabla_a X^0.
\eeq
The form $\nabla_a X^0$ depends explicitly on the $X$ field, and hence is not allowed in our
constructions.  However, the unit normal $u_a$ can be constructed using only the surface $\Sigma$
and the metric, and hence is independent of the $X$ fields.  This then implies that $N$ also depends
on the $X$ fields, and so the lapse function cannot explicitly be used in constructing ambiguity terms.

\subsection{$L$ ambiguity}
The first ambiguity corresponds to adding an exact form $d\alpha$ to the Lagrangian.   This 
does not affect the equations of motion; however, its variation now contributes to $\theta$.  
The following changes occur from adding this term to the Lagrangian:
\begin{subequations}
\begin{align}
L&\rightarrow L + d\alpha \\
\theta&\rightarrow \theta+\del\alpha\\
J_\xi&\rightarrow J_\xi + d i_\xi \alpha \\
Q_\xi&\rightarrow Q_\xi + i_\xi \alpha. \label{eqn:QLambig}
\end{align} 
\end{subequations}
Note that since $\theta$ changes by an $\SSS$-exact form, the symplectic current $\omega$
is unaffected.  Incorporating these changes into the definition of the symplectic potential 
 (\ref{eqn:ThS}) changes $\Theta$ by
 \beq
 \Th \rightarrow \Th + \int_\Sigma(\del\alpha+i_\dx d\alpha)+\int_{\partial\Sigma} i_\dx \alpha
 = \Th + \del\int_\Sigma\alpha.
 \eeq
 We point out that the new term annihilates infinitesimal
 diffeomorphisms $I_{\hat \xi}$, so that $\Th$ remains fully diffeomorphism-invariant.
 Since $\Theta$ changes by an $\SSS$-exact form, the symplectic form $\Omega =\del\Th$ 
receives no change from this type of ambiguity, which can also be checked by tracking
the changes of all quantities in (\ref{eqn:OmS}).  
Given that only $\Omega$, and not $\Theta$, is needed in the construction of the phase space,
this ambiguity in $L$ has no effect on the phase space.  However, it
has some relevance to the surface symmetry algebra discussed in section \ref{sec:ssa}.  
The generators of this algebra are given by the Noether charge, and for surface symmetries that 
move $\partial \Sigma$ (the ``surface translations''), this ambiguity would appear to 
have an effect.  However, as discussed in subsection \ref{sec:trans}, once the appropriate
boundary terms are included in the generators, 
the result is independent of this ambiguity.  The 
 form of the generator does motivate a natural prescription for fixing the ambiguity 
 such that the Lagrangian has 
a well-defined variational principle, so that it is completely stationary on-shell, as opposed to 
being stationary up to boundary contributions.

\subsection{$\theta$ ambiguity} \label{sec:thambig}
The second ambiguity comes from the freedom to add an exact form $d\beta$ to $\theta$,
since doing so does not affect its defining equation (\ref{eqn:dL}).  Here, $\beta\equiv\beta[\phi;
\del\phi]$ is a spacetime $(d-2)$-form and a one-form on $\SSS$.  
The changes that arise from this addition are
\begin{subequations}
\begin{align}
\theta&\rightarrow \theta + d\beta \label{eqn:beta}\\
\omega&\rightarrow \omega + d \del \beta\\
J_\xi &\rightarrow J_\xi + d I_{\hat\xi}\beta \\
Q_\xi & \rightarrow Q_\xi + I_{\hat\xi}\beta. \label{eqn:Q+Ixib}
\end{align}
\end{subequations}
Under these transformations, the symplectic potential (\ref{eqn:ThS}) changes to
\beq
\Th\rightarrow \Th +\int_{\partial\Sigma} (\beta + I_{\hdx}\beta).
\eeq

Hence, the symplectic potential is modified by an arbitrary boundary term $\beta$, 
accompanied by $I_{\hdx}\beta$ that ensures that $\Th$ retains degenerate directions
along linearized diffeomorphisms.  Unlike the $L$ ambiguity, this modification is not $\SSS$-exact,
and changes the boundary terms in the symplectic form,
\beq
\Om\rightarrow\Om +\int_{\partial\Sigma} (\del\beta+\del I_{\hdx}\beta+\lie_{\dx} \beta
+\lie_\dx I_{\hdx}\beta).
\eeq
Because $\beta$ can in principle involve arbitrarily many derivatives of $\del\phi$, its presence
can cause $\Omega$ to depend on second or higher derivatives of $\dx^a$ on the boundary.  
This affects which parts of $\dx^a$ correspond to degenerate directions, and will lead to 
different numbers of boundary degrees of freedom in the reduced phase space.  As discussed
in section \ref{sec:ssa}, this ambiguity can also be used to reduce the surface symmetry 
algebra to a subalgebra.  

Give that $\beta$  contributes to $\Th$ and $\Om$ only at the boundary, it can involve tensors 
associated with the surface $\partial\Sigma$ that do not 
correspond
to spacetime-covariant tensors, such as the extrinsic curvature.  This allows the Dong entropy
\cite{Dong2014, Camps2013, Miao2014}, 
which differs from the Wald entropy \cite{Wald1993a, Iyer1994a} by extrinsic curvature
terms, to 
be viewed as a Noether charge with a specific choice of ambiguity terms.  This is the
point of view 
advocated for in \cite{Wall2015}, where the ambiguity was resolved by requiring that the 
entropy functional derived from the resultant Noether charge satisfy a linearized second law.  
In general, fixing the ambiguity requires some additional input, motivated by the particular
application at hand.

\subsection{$Q_\xi$ ambiguity}
The final ambiguity is the ability to shift $Q_\xi$ by a closed form $\gamma$, with $d\gamma=0$.
Since $Q_\xi$ depends linearly on $\xi^a$ and its derivatives, 
$\gamma$ should be chosen 
to also satisfy this requirement.  If $\gamma$ is identically closed for all $\xi^a$, it
then follows that it must be exact, $\gamma = d\nu$ \cite{Wald1990a}.  
Its integral over the closed surface
$\partial\Sigma$ then vanishes, so that it has no effect on $\Theta$ or $\Omega$.

\section{Surface symmetry algebra}\label{sec:ssa}
The extended phase space constructed in section \ref{sec:eps} contains new edge mode
fields $\dx^a$  on the boundary of the local subregion, whose presence is required 
in order to
have a gauge-invariant symplectic form.  Associated with the edge modes are a new
class of transformations that leave the symplectic form and the equations of motion 
invariant.  These new transformations
comprise the surface symmetry algebra.  This algebra 
plays an important role  in the quantum theory 
when describing the edge mode contribution to the entanglement entropy, thus it is necessary
to identify the algebra and its canonical generators.  

As discussed in \cite{Donnelly2016}, the surface symmetries coincide with diffeomorphisms 
in the preimage space, $Z: \mathbb{R}^d\rightarrow \mathbb{R}^d$, where $\mathbb{R}^d
\supset X^{-1}(M)$. 
These leave the spacetime fields $\phi$ unchanged, but transform
the $X$ fields by $X\rightarrow X\circ Z$.  This also transforms the pulled back fields 
$X^*\phi\rightarrow Z^* X^*\phi$, and due to the diffeomorphism invariance of the field
equations, the pulled back fields still define solutions.  These transformations therefore
comprise a set of symmetries for the dynamics in the local subregion.  
Infinitesimally, these transformations are generated
by vector fields $w^a$ on $\mathbb{R}^d$.  Analogous to 
vector fields defined on $M$, $w^a$ defines
a vector $\hat{w}$ on $\SSS$, whose action on the pulled back fields $X^*\phi$ is given by the 
Lie derivative,
\beq
L_{\hat{w}} X^*\phi = \lie_w X^*\phi = X^*\lie_{(X^{-1})^*w}\phi,
\eeq
while its action on $\phi$ is trivial, $L_{\hat w}\phi = 0$.  On the other hand, we may apply the 
pullback formula (\ref{eqn:LphX*a}) to this equation to derive
\begin{align}
X^*\lie_W\phi = X^* I_{\hat w} \lie_\dx \phi,
\end{align}
where $W^a = ({X^{-1}})^*w^a$.  The contractions of the vector $\hat{w}$ with the basic $\SSS$ 
one-forms are therefore
\beq
I_{\hat w} \dar^a = W^a,\qquad I_{\hat w}\del\phi=0.
\eeq
We also will assume that $w^a$ is independent of the solution, so that $\del w^a=0$.  
Writing this as $0=\del X^* W^a$, and applying the pullback formula (\ref{eqn:LphX*a}), one 
finds
\beq\label{eqn:dWa}
\del W^a = -\lie_{\dx} W^a.
\eeq 

In order for the transformation to be a symmetry of the phase space, it must generate a 
Hamiltonian flow.  This means that $I_{\hat w} \Omega$ is exact, and determines the Hamiltonian
$H_{\hat w}$
for the flow via $\del H_{\hat w} = - I_{\hat w}\Om$.  The contraction with the symplectic form
can be computed straightforwardly from (\ref{eqn:OmS}) by first using the decomposition 
(\ref{eqn:dQdX}) for $\del Q_\dx$.  Then
\begin{align}
I_{\hat w}\Om &= \int_{\partial\Sigma}(-\qo_W-Q_{[W,\dx]} -\lie_\dx Q_W +\lie_W Q_\dx+i_W\theta
+i_W i_\dx L) \\
&=-\del \int_{\partial\Sigma}Q_W + \int_{\partial\Sigma} i_W(\theta+I_{\hdx}\theta). 
\label{eqn:diQW}
\end{align}
The first three terms of the first line combine into the first term in the second line, using  formula
(\ref{eqn:dWa}) for $\del W^a$, formula (\ref{eqn:dQdX}) for $\del Q_W$, and recalling that 
the integral involves an implicit pullback by $X^*$, so that $\del \int_{\partial\Sigma} Q_W
 = \int_{\partial \Sigma} (\del Q_W +\lie_\dx Q_W)$.  
 
It is immediately apparent that if the second integral in (\ref{eqn:diQW}) vanishes, the flow 
is Hamiltonian.  This  occurs if $W^a$ is tangent to $\partial \Sigma$ or vanishing at 
$\partial\Sigma$, and hence defines a mapping of the surface into itself.  If $W^a$ is tangential,
it generates a diffeomorphism $\partial\Sigma$, while vector fields that vanish on $\partial \Sigma$
generate transformations of the normal bundle to the surface while holding all points on the 
surface fixed.  These transformations were respectively called surface diffeomorphisms and 
surface boosts in \cite{Donnelly2016}.  The remaining transformations consist of the 
surface translations, where 
$W^a$ has components normal to the surface, and the second integral in (\ref{eqn:diQW})
does not vanish.  In general, this term does not give a Hamiltonian flow, except when the 
fields satisfy certain boundary conditions.  We will briefly discuss the surface translations
in subsection \ref{sec:trans}, where we show that they can give rise to central charges in the 
surface symmetry algebra.

Returning to the surface-preserving transformations, we find that the Hamiltonian is 
given by the Noether charge integrated over the boundary,
\beq
H_{\hat w} = \int_{\partial \Sigma} Q_W.
\eeq
The surface symmetry algebra is generated through the Poisson bracket of the Hamiltonians
for all possible surface-preserving vectors.  The Poisson bracket is given by
\beq
\{ H_{\hat w}, H_{\hat v}\} = I_{\hat w} I_{\hat v}\Omega = -I_{\hat w} \del\int_{\partial \Sigma} Q_V
=-I_{\hat w}\int_{\partial \Sigma}(\qo_V+Q_{\del V} + \lie_\dx Q_V) = \int_{\partial\Sigma}Q_{[W,V]},
\eeq
where the last equality uses equation (\ref{eqn:dWa}) applied to $\del V^a$ and that 
$\int_{\partial\Sigma}\lie_WQ_V = \int_{\partial\Sigma} i_W dQ_V$
vanishes when integrated over the surface since $W^a$ is parallel to $\partial\Sigma$.
This shows that the algebra generated by the Poisson bracket is compatible with the 
Lie algebra of surface preserving vector fields,
\beq
\{H_{\hat w}, H_{\hat v}\} = H_{{[w,v]}\,\hat{}},
\eeq
without the appearance of any central charges, i.e.\ the map $w^a\mapsto H_{\hat w}$ 
is a Lie algebra homomorphism.  
Note that the algebra of surface-preserving vector fields is much larger than the 
surface symmetry algebra.  This is because the  generators of surface symmetries
depend only on the values of the vector field and its derivative at $\partial \Sigma$.  Vector fields
that die off sufficiently quickly near $\partial\Sigma$ correspond to vanishing Hamiltonians.  
The transformations they induce on $\SSS$ are pure gauge, and they drop out after passing 
to the reduced phase space.  

To identify the surface symmetry algebra, it is useful to first describe the larger algebra of 
surface-preserving diffeomorphisms, which contains the surface symmetries as a subalgebra. 
It takes the form of a semidirect product, $\text{Diff}(\partial\Sigma)\ltimes\ddiff$
where 
$\text{Diff}(\partial \Sigma)$ is the diffeomorphism group of $\partial\Sigma$, and $\ddiff$ is 
the normal subgroup of diffeomorphisms that fix all points on $\partial\Sigma$.\footnote{
``$\text{Dir}$'' stands for ``Dirichlet,'' since these are the diffeomorphisms that would be 
consistent with fixed, Dirichlet boundary conditions at $\partial\Sigma$.}  $\ddiff$ is 
generated by vector fields $W^a$ that vanish on $\partial\Sigma$, and it is a normal subgroup 
because the vanishing property is preserved
under commutation with all surface-preserving vector fields:
\beq
[W,V]^a\big|_{\partial\Sigma} = (W^b\partial_bV^a-V^b\partial_bW^a)\big|_{\partial\Sigma}=0,
\eeq
where the first term vanishes since $W^b$ vanishes at $\partial\Sigma$, and the second term
vanishes because $V^b$ is parallel to $\partial\Sigma$, and $W^a$ is zero everywhere along 
the surface.  A general surface preserving vector field can then be expressed as 
\beq
W^a = W^a_\parallel+ W^a_0,
\eeq
where $W^a_0$ vanishes on $\partial\Sigma$ and $W^a_\parallel$ is tangent to $\partial \Sigma$.  
Note that this decomposition is not canonical; away from $\partial\Sigma$ there is some freedom
in specifying which components of the vector field correspond to the tangential direction.  
However, given any such
choice, it is clear that 
if $W^a_\parallel$ is nonvanishing at $\partial\Sigma$, then it will be nonzero in a 
neighborhood of $\partial\Sigma$, and hence the parallel vector fields  
act nontrivially on the $V_0^a$ component of 
 other vector fields.  Finally, the commutator of two purely parallel vector fields $[W_\parallel, 
 V_\parallel]$ will remain purely parallel, since they are tangent to an integral submanifold.  
The map $W^a\mapsto W^a_\parallel$ is therefore a homomorphism from the surface-preserving
diffeomorphisms onto $\text{Diff}(\partial\Sigma)$, with kernel  $\ddiff$.  This
establishes that the group of surface-preserving diffeomorphisms is $\text{Diff}(\partial\Sigma)
\ltimes \ddiff$.

The surface symmetry algebra is represented as a subalgebra of $\text{Diff}(\partial\Sigma)
\ltimes \ddiff$.  The Hamiltonian for a surface-preserving vector field is determined by the 
Noether charge $Q_W$, which depends only on the value of $W^a$ and its first derivative at 
$\partial\Sigma$.  Hamiltonians for vector fields that are nonvanishing at $\partial\Sigma$ provide
a faithful representation of the $\text{Diff}(\partial\Sigma)$ algebra; however, the vanishing
vector fields only represent a subalgebra of $\ddiff$. To determine it, note that only the first
derivative of $W^a$ contributes to the Noether charge, and its tangential derivative vanishes.  
Letting $x^i$, $i=0,1$, 
represent coordinates in the normal directions that vanish on $\partial\Sigma$, 
the components of the vector field may be expressed $W^\mu = x^i W\indices{_i^\mu}
+\mathcal{O}(x^2)$, $\mu=0,\ldots,d-1$, 
and the $\mathcal{O}(x^2)$ 
terms are determined by the second derivatives, which do not contribute
to the Noether charge.  Then
the  commutator of two vectors is
\beq
[W,V]^\mu = x^i(W\indices{_i^j}V\indices{_j^\mu}-V\indices{_i^j}W\indices{_j^\mu})+
\mathcal{O}(x^2),
\eeq
which is seen to be determined by the matrix commutator of $W\indices{_i^\mu}$ and 
$V\indices{_j^\nu}$, by allowing the $i,j$ indices to run over $0,\ldots, d-1$, setting all entries
with $i,j>1$ to zero.  

This algebra gives a copy of $SL(2,\mathbb{R})\ltimes \mathbb{R}^{2\cdot(d-2)}$ for each
point on $\partial \Sigma$.  The abelian normal subgroup $\mathbb{R}^{2\cdot(d-2)}$ is 
generated by vectors for which the $\mu$ index in $W\indices{_i^\mu}$ is tangential, i.e.\
$W\indices{_i^j}\equiv W\indices{_i^\mu}\nabla_\mu x^j=0$.
These vectors represent shearing transformations of the normal bundle: they 
generate flows that vanish on $\partial\Sigma$, and are parallel to $\partial \Sigma$ 
away from the surface.   By specifying a normal direction,
one obtains a homomorphism sending $W\indices{_i^\mu}$ 
to its purely normal part, $W\indices{_i^j}$.  The fact that only the traceless part of $\nabla_aW^b$
contributes to the Noether charge, which follows from the antisymmetry of $E^{abcd}$ from 
equation (\ref{eqn:Qxi}) in $c$ and $d$, translates to the requirement that 
$W\indices{_i^j}$ be traceless when $W^a$ vanishes on $\partial\Sigma$.  This means that 
the $2\times 2$ matrices $W\indices{_i^j}$ generate an $SL(2,\mathbb{R})$ algebra.  The 
generators $V\indices{_i^\mu}$  of $\mathbb{R}^{2\cdot(d-2)}$ transform as a collection of 
$(d-2)$ vectors 
under the $SL(2,\mathbb{R})$ algebra, verifying the semidirect product structure 
$SL(2,\mathbb{R})\ltimes \mathbb{R}^{2\cdot(d-2)}$ for the vector fields vanishing at 
$\partial\Sigma$.  Under diffeomorphisms of $\partial\Sigma$, $V\indices{_i^\mu}$ transforms
as a pair of vectors; hence,
the full surface symmetry algebra is  $\text{Diff}(\partial\Sigma)
\ltimes\left(SL(2,\mathbb{R})\ltimes \mathbb{R}^{2\cdot(d-2)}\right)^{\partial\Sigma}$.

The extra factor of $\mathbb{R}^{2\cdot(d-2)}$ is a novel feature of this analysis, appearing 
for generic higher curvature theories, but not for general relativity \cite{Donnelly2016}.  
Its presence or absence 
is explained by the particular structure of $E^{abcd}$, the variation of the Lagrangian scalar with
respect to $R_{abcd}$.  When $E^{abcd}$ is determined by its trace, i.e., equal to 
$\frac{E}{d(d-1)} (g^{ac}g^{bd}-g^{ad}g^{bc})$ with $E$ a scalar, the $\mathbb{R}^{2\cdot(d-2)}$ 
transformations are pure gauge.  The Noether charge for a vector field vanishing
at the surface evaluates to\footnote{The binormal is defined to be $n_{ab} = 2u_{[a}n_{b]}$ where
$u_a$ is the timelike unit normal and $n_a$ is the inward-pointing
spacelike unit normal.  The spacetime
volume form at $\partial \Sigma$ is then $\ep_{ab}\big|_{\partial\Sigma} = -n_{ab}\wedge\mu$.} 
\beq
Q_W\big|_{\partial\Sigma} =\mu\, n_{ab}E\indices{^a^b^c_d}\nabla_cW^d
= \mu \frac{E}{d(d-1)} n\indices{^c_d}\nabla_c W^d,
\eeq
where $\mu$ is the volume form on $\partial\Sigma$ and $n_{ab}$ is the binormal;
 $n\indices{^c_d}$ projects out the tangential component in $\nabla_cW^d$, leaving only
the $SL(2,\mathbb{R})$ transformations as physical symmetries.  A particular class of theories
in which this occurs are $f(R)$ theories (which include general relativity), 
where the Lagrangian is a function of the Ricci scalar,
and $E^{abcd} = \frac12f'(R)(g^{ac}g^{bd}-g^{ad}g^{bc})$.  In more general theories, however,
$n_{ab}E\indices{^a^b^c_d}$ will have a tangential component on the $d$ index, and the algebra
enlarges to include the $\mathbb{R}^{2\cdot(d-2)}$ tranformations.

Curiously, there always exists a choice of ambiguity terms, discussed in subsection 
\ref{sec:thambig},
that eliminates the $\mathbb{R}^{2\cdot(d-2)}$ symmetries.  Namely, 
the symplectic potential current $\theta$ can be modified 
as in equation (\ref{eqn:beta}), with $\beta$ 
chosen to be 
\beq \label{eqn:bmod}
\beta = \ep_{ab}E^{abed}s\indices{_e^c} \del g_{cd},
\eeq
and $s\indices{_e^c} = -u_e u^c + n_e n^c$ is the projector onto the normal bundle
of $\partial\Sigma$.  Note that the explicit use of  normal vectors to $\partial \Sigma$
makes this $\beta$ not spacetime-covariant.
This is nevertheless in line with the broader set of allowed ambiguity terms discussed above.  
From equation (\ref{eqn:Q+Ixib}), this term changes 
the Noether charge of a vector vanishing at $\partial\Sigma$ to 
\beq
Q_W\big|_{\partial\Sigma}=\mu \,n_{ab}\left(E\indices{^a^b^c_d}-E\indices{^a^b^e_d}s\indices{_e^c}
-E\indices{^a^b^e^c}s\indices{_e_d}  \right)\nabla_c W^d.
\eeq
The additional terms involving $s\indices{_e^c}$ drop out when contracted with the normal 
component on the $d$ index of $\nabla_c W^d$; however, on the tangential component the
additional terms
cancel against the first term.  This choice of ambiguity thus reduces the 
surface symmetry algebra to coincide with the algebra for general relativity, 
$\text{Diff}(\partial\Sigma)\ltimes SL(2,\mathbb{R})^{\partial\Sigma}$.

Whether or not to use this choice of $\beta$ depends on the application at hand, and it is 
unclear at the moment how exactly $\beta$ should be fixed when trying to characterize the 
edge mode contribution to the entanglement entropy of a subregion.  The above choice is 
natural in the sense that it gives the same surface symmetry algebra for any 
diffeomorphism-invariant theory.  This would mean that the surface symmetry algebra is 
determined by the gauge group of the theory, while the Hamiltonians for the symmetry generators 
change depending on the specific dynamical theory under consideration.  Note also that there
are additional ambiguity terms that could be added, some of which enlarge the symmetry algebra
by introducing dependence on higher derivatives of the vector field.  Determining how to fix 
the ambiguity remains an important open problem for the extended phase space program.

\subsection{Surface translations} \label{sec:trans}
While the surface-preserving transformations are present for generic surfaces, in situations 
where the fields satisfy certain boundary conditions at $\partial\Sigma$, the surface-symmetry
algebra can enhance to include surface translations.  These are generated by vector fields
that contain a normal component to $\partial\Sigma$ on the surface.  For such a vector field,
the second integral in (\ref{eqn:diQW}) does not vanish, so for this transformation to be 
Hamiltonian, this integral must be an exact $\SSS$ form.  To understand when this can
occur, it is useful to first rewrite the integral in terms of pulled back fields on $\partial\sigma$,
the preimage of $\partial\Sigma$ under the $X$ map:
\beq
\int_{\partial\Sigma} i_W(\theta+I_{\hdx}\theta) = 
\int_{\partial\sigma}X^* i_W\theta[\phi;\del\phi +\lie_\dx\phi] 
=\int_{\partial\sigma} 
i_w\theta[X^*\phi; \del X^*\phi].
\eeq
Since $\delta w^a=0$, it is clear from this last expression that the flow will be Hamiltonian 
only if at the boundary, $\theta$ is exact when contracted with $w^a$,
\beq \label{eqn:iwthds}
i_w\theta[X^*\phi;\del X^*\phi]\big|_{\partial\sigma} = i_w\delta X^*B,
\eeq
where $B[\phi]$ is some functional of the fields, possibly involving structures
defined only at $\partial\Sigma$ such as the extrinsic curvature.  When this condition is 
satisfied, the second integral in (\ref{eqn:diQW}) simply becomes $\del\int_{\partial\Sigma} i_W B$,
and so the full Hamiltonian for an arbitrary vector field $w^a$ is
\beq \label{eqn:Hhw}
H_{\hat w}=\int_{\partial\Sigma} \left(Q_W-i_WB\right).
\eeq

Next we compute the algebra of the surface symmetry generators under the Poisson bracket.  
It is worth noting first that by contracting equation (\ref{eqn:iwthds}) with $I_{\hat v}$, 
we find that the $B$ functional satisfies
\beq
i_W \lie_V B\big|_{\partial\Sigma} = i_W I_{\hat V}\theta = i_W(dQ_V+i_VL).
\eeq
With this, the Poisson bracket is given by
\begin{align}
\{H_{\hat w}, H_{\hat v} \} &= -I_{\hat w} \del\int_{\partial\Sigma} \left(Q_V-i_V B\right)   \nonumber \\
&= \int_{\partial\Sigma} \left(-I_{\hat w}\del Q_V - \lie_W Q_V+I_{\hat w} i_{\del V} B + \lie_W
i_V B  \right) \nonumber \\
&=\int_{\partial\Sigma}\left( Q_{[W,V]}-i_{[W,V]}B\right) + \int_{\partial\Sigma}i_W\left(-dQ_V
+\lie_V B - i_V dB\right) \nonumber \\
&= H_{[w,v]\,\hat{}} + \int_{\partial\Sigma} i_W i_V(L-dB). \label{eqn:Hwvh}
\end{align}
Hence, the commutator algebra of the vector fields $w^a$ is represented by the algebra
provided by the Poisson bracket, except when both vector fields have normal components 
at the surface, in which case the second term in (\ref{eqn:Hwvh}) gives a modification.
In fact, the quantities
\beq\label{eqn:Kwhvh}
K[\hat w, \hat v]\equiv \int_{\partial\Sigma} i_W i_V (L-dB)
\eeq
provide a central extension of the algebra, which is verified by showing that 
they are locally constant on the phase space, and hence commute with all generators.  
The exterior derivative is
\begin{align}
\del K[\hat w, \hat v] &= \int_{\partial\Sigma} \big[\del i_W i_V(L-dB) + \lie_\dx i_W i_V (L-dB)\big] 
\nonumber\\
&= \int_{\partial\Sigma} i_W i_V (\del L-d\del B).
\end{align}
On shell, we have $\del L =  d\theta$, and from (\ref{eqn:iwthds}) we can argue
that the replacement $ i_W i_Vd\del B\rightarrow i_W i_Vd\theta$ is valid at $\partial\Sigma$.  
Hence, the above variation vanishes, and $K[\hat w, \hat v]$ indeed defines a central extension
of the algebra.  

The modification that $B$ makes to the symmetry generators takes the same form as a Noether
charge ambiguity arising from changing the Lagrangian $L\rightarrow L+d\alpha$, with $\alpha=-B$.
Using the 
modified Lagrangian
$L-dB$, the potential current changes to $\theta-\del B$.  The boundary condition
(\ref{eqn:iwthds}) then implies that the terms involving $\theta$ in (\ref{eqn:diQW}) vanish.  
The symmetry generators are simply given by the integrated Noether charge, which is modified to
modified to $Q_W\rightarrow Q_W-i_W B$ by the ambiguity.  Hence, the generators $H_{\hat w}$ 
are the same as in (\ref{eqn:Hhw}), and their Poisson brackets  still involve the central 
charges $K[\hat w, \hat v]$. Finally, note that the constancy of the central charges 
requires the variation of the modified Lagrangian $L-dB$ be zero
when evaluated on 
$\partial\Sigma$.  Requiring that variations of the Lagrangian have no boundary term on
shell generally determines the boundary conditions for the theory.  The same is true
here: a choice of $B$ satisfying (\ref{eqn:iwthds}) can generally only be found if the fields 
obey certain boundary conditions, and different boundary conditions lead to different choices
for $B$.  

The surface translations 
can be parameterized
by normal vector fields $W^i$ defined on $\partial\Sigma$.  Assuming 
$\partial_i W^j=0$ in some coordinate system, where $i,j$ are normal indices, 
we can work out their commutation relations
with generators of the rest of the algebra:
\begin{align}
[W^i, V^j] &=0 \label{eqn:WV}\\
[W^i, x^j V\indices{_j^k}] &= W^i V\indices{_i^k} \\
[W^i, V^A] &= V^A\partial_A W^i\\
[W^i, x^j V\indices{_j^A}] &= W^i V\indices{_i^A}- x^j V\indices{_j^A}\partial_A W^i, 
\label{eqn:WxV}
\end{align}
where $A$ denotes a tangential index.  The first relation shows that the new generators commute
among themselves (although the corresponding Poisson bracket is equal to the central charge
$K[\hat w, \hat v]$), while the second and third show that $W^i$ transforms as a vector under
$SL(2,\mathbb{R})$ and as a scalar under $\text{Diff}(\partial \Sigma)$.  If the Noether charge
ambiguity is chosen as in equation (\ref{eqn:bmod}) 
so that the normal shearing generators $x^j V\indices{_j^A}$  
drop out of the algebra, the resulting surface symmetry algebra is $\text{Diff}(\partial\Sigma)\ltimes
\left(SL(2,\mathbb{R})\ltimes \mathbb{R}^2\right)^{\partial\Sigma}$.
However, if the normal shearing transformations are retained, equation (\ref{eqn:WxV}) shows
that the surface translations are no longer a normal subgroup, since the commutator gives 
rise to generators of $\text{Diff}(\partial\Sigma)$ and $SL(2,\mathbb{R})^{\partial\Sigma}$.  
In this case, the full surface symmetry algebra is simple.

The above analysis was carried out assuming that all normal vectors generate a surface 
symmetry.  In practice, equation (\ref{eqn:iwthds}) may only be obeyed for some 
specifically chosen normal vectors \cite{Brown1986a}.  The resulting 
algebra will then be a subalgebra of the generic case considered in this section.

\section{Discussion} \label{sec:disc}

Building on the results of \cite{Donnelly2016}, this paper has described a general procedure 
for constructing the extended phase space in a 
diffeo\-morphism-invariant theory for a local subregion.  
The integral of the symplectic current for the unextended theory fails to be degenerate for 
diffeomorphisms that act at the boundary, and this necessitates the introduction of new fields,
$X$, to ensure degeneracy.  These fields can be thought of as defining a coordinate system
for the local subregion, and the extended solution space consists of fields satisfying 
the equations of motion in all possible coordinate systems parameterized by $X$.  
While the $X$ fields do not satisfy dynamical equations themselves, it was shown in 
section \ref{sec:eps} 
 that their variations contribute to the symplectic form through the boundary
integral  in equation (\ref{eqn:OmS}).  

There are a few novel features of the extended phase space for arbitrary diffeomorphism-invariant
theories that do not arise in vacuum
general relativity with zero cosmological constant.  First, in any
theory whose Lagrangian does not vanish on-shell, the symplectic potential $\Theta$ is not
a single-valued one form on the reduced phase space $\PP$.  This is due to the bulk integral
of the Lagrangian that appears in  equation (\ref{eqn:ThS}), along with the fact that variations 
for which $\dx^a$ has support only away from the boundary $\partial\Sigma$ are 
degenerate directions of the extended symplectic form, (\ref{eqn:OmS}).  
Because of this, $\Om$ fails to be exact, despite satisfying $\del \Om=0$.  Investigating the 
consequences of this nontrivial cohomology for $\PP$ remains an interesting topic for 
future work.  

Another new result comes from the form of the surface symmetry algebra.  
As in general relativity, any phase space transformation generated by $\hat w$ 
for which $W^a\equiv I_{\hat w} \dx^a$ is tangential  at $\partial \Sigma$
is Hamiltonian. These generate the group $\text{Diff}(\partial\Sigma)\ltimes \ddiff$ of 
surface-preserving diffeomorphisms, but only a subgroup is represented on the phase space.  
This subgroup was found in section \ref{sec:ssa} 
to be $\text{Diff}(\partial\Sigma)\ltimes\left(SL(2,\mathbb{R})\ltimes
\mathbb{R}^{2\cdot(d-2)} \right)^{\partial\Sigma}$, which is larger than the surface symmetry
group $\text{Diff}(\partial\Sigma)\ltimes SL(2,\mathbb{R})^{\partial\Sigma}$ found in
\cite{Donnelly2016} for general relativity.  The additional abelian factor $\mathbb{R}^{2\cdot(d-2)}$
arises generically; however, 
it is not present in 
$f(R)$ theories, in which the tensor $E^{abcd}$ is constructed solely
from the metric and scalars.  We also noted that for any theory, there exists a choice
(\ref{eqn:bmod})
of ambiguity terms that can be added to $\theta$, with the effect of eliminating the 
$\mathbb{R}^{2\cdot(d-2)}$ factor of the surface symmetry algebra.  

The inclusion of  surface 
translations into the surface symmetry algebra was discussed in section \ref{sec:trans}.  
This requires the existence of a $(d-1)$-form $B$ satisfying the relation
(\ref{eqn:iwthds}) for at least some vector fields that are normal to the boundary.  
If such a form can be found, the surface translations are generated by the Hamiltonians
(\ref{eqn:Hhw}).
Interestingly, the Poisson brackets of these Hamiltonians acquire central charges
given by (\ref{eqn:Kwhvh}), which depend on the on-shell value of the modified Lagrangian
$L-dB$ at $\partial\Sigma$.  
Such central charges are a common occurrence in surface symmetry algebras that include
surface translations \cite{Brown1986a, Carlip1999, Silva2002, Barnich2002, 
Compere2007,Carlip2011, Freidel2017}.
In general, the existence of $B$ requires that the fields satisfy boundary conditions at 
$\partial\Sigma$.  An important topic for future work would be to classify which boundary
conditions the fields must satisfy in order for $B$ to exist.  For example, with 
Dirichlet boundary conditions where
the field values are specified at $\partial\Sigma$, $B$ is given by the Gibbons-Hawking
boundary term, constructed from the trace of the extrinsic curvature in the normal
direction \cite{Gibbons1977}.   However, 
such boundary conditions are quite restrictive on the dynamics.  For a local subsystem in which
$\partial\Sigma$ simply represents a partition of a spatial slice, one would not expect Dirichlet
conditions to be compatible with all solutions of the theory.  An alternative approach would be
to impose conditions that specify the location of the surface in a diffeomorphism-invariant manner,
without placing any restriction on the dynamics.  One example is requiring that the surface
extremize its area or some other entropy functional, as is common in holographic entropy 
calculations \cite{Ryu:2006bv, Hubeny2007, Dong2014, Camps2013, Miao2014, Dong2017}.  
Since extremal surfaces exist in generic solutions, these boundary conditions put no
dynamical restrictions on the theory, but rather restrict where the surface $\partial\Sigma$ lies.  

The effects of JKM ambiguity terms in the extended phase space 
construction were discussed in section \ref{sec:jkm}.  It was noted that the $B$ form that appears
when analyzing the surface translations could be interpreted as a Lagrangian ambiguity,
 $L\rightarrow L-dB$.  Note that this type of ambiguity does not affect the symplectic 
form (\ref{eqn:OmS}), and, as a consequence, the generators of the surface 
symmetries do not depend on this replacement.  In fact, the generators (\ref{eqn:Hhw}) are 
invariant with respect to additional changes to the Lagrangian $L\rightarrow L+d\alpha$,
since such a change shifts the Noether charge $Q_W\rightarrow Q_W + i_W\alpha$, but
also induces the change $B\rightarrow B+\alpha$.  An ambiguity that does affect the phase
space is the shift freedom in the symplectic potential current, $\theta\rightarrow\theta+d\beta$.
We noted that certain choices of $\beta$ can change the number of edge mode degrees of 
freedom, and also can affect the surface symmetry algebra.  
In the future, we would like to understand how this ambiguity should be fixed.  One idea 
would be to use the ambiguity to ensure some $B$ can be found satisfying equation 
(\ref{eqn:iwthds}).  In this case, the ambiguity is fixed as an integrability condition
for $\theta$.  Such an approach seems related to the ideas of \cite{Wall2015} in which
the ambiguity was chosen to give an entropy functional satisfying a linearized second law.  
Another approach discussed in \cite{Miao2014, Miao2015b, Camps2016, Dong2017} fixes
the ambiguity through the choice of metric splittings that arise when performing the
replica trick in the computation of holographic entanglement entropy.  

As discussed in the introduction, one of the main motivations for 
constructing the extended phase space is to understand entanglement entropy
in diffeomorphism-invariant theories \cite{Donnelly2016}.  
The Hilbert space for such a theory does not factorize
across an entangling surface due to the constraints.  However, one can instead construct an 
extended Hilbert space for a local subregion $\Sigma$ as a quantization of the 
extended phase space constructed above.  This extended Hilbert space will contain edge mode
degrees of freedom that transform in representations of the surface symmetry algebra.  
A similar extended Hilbert space can be constructed for the complementary region $\bar{\Sigma}$,
whose edge modes and surface symmetries will match those associated with $\Sigma$.  
The physical Hilbert space for $\Sigma\cup \bar\Sigma$ is given by the so-called entangling
product of the two extended Hilbert spaces, which is  the tensor product modded out by the 
action of the surface symmetry algebra.  One then finds that the density matrix associated with
$\Sigma$ splits into a sum over superselection sectors, labelled by the representations of the 
surface symmetry group.  

This block diagonal form of the density matrix leads to a 
von Neumann entropy that is the sum of three types of terms,
\beq
S = \sum_i \left( p_i S_i  - p_i \log p_i + p_i \log \dim R_i\right),
\eeq
where the sum is over the representations $R_i$ of the surface symmetry group, $p_i$ give the 
probability of being in a given representation, and $S_i$ is the von Neumann entropy within
each superselection sector.  The first term represents the average entropy of the interior
degrees of freedom, while the second term is a classical Shannon entropy coming from 
uncertainty in the surface symmetry representation corresponding to the state.  The last 
term arises from entanglement between the edge modes themselves, and is only present for a 
nonabelian surface symmetry algebra \cite{Donnelly2012a, Donnelly2014a}.
The dimension of the representation has some expression in terms of the Casimirs of the 
group, and hence this term will take the form of an expectation value of local operators
at the entangling surface.  It is conjectured that this term provides a statistical
interpretation for the Wald-like contributions in the generalized entropy, $S_\text{gen}
= S_\text{Wald-like} + S_\text{out}$ \cite{Donnelly2016}.  Put another way, given a UV 
completion for the quantum gravitational theory, the edge modes keep track of the
 entanglement between the UV modes that are in a fixed state, corresponding
to the low energy ``code subspace'' \cite{Lin2017, Harlow2016}.  

On reason for considering the extended phase space in the context of entanglement
entropy comes from issues of divergences in entanglement entropy.  These divergences
arise generically in quantum field theories, and a regulation prescription is needed 
in order to get a finite result.  A common regulator for Yang-Mills theories is a lattice 
\cite{Casini2014a, Donnelly2012a, Donnelly2014a},
which preserves the gauge invariance of the theory.  Unfortunately, a lattice 
breaks diffeomorphism invariance, which can be problematic when 
using it as a regulator for gravitational
theories (see \cite{Hamber2009} for a review of the lattice approach to quantum
gravity).  The extended phase space provides a continuum description of the edge modes that
respects diffeomorphism invariance.  As such, it should be amenable to finding a regulation
prescription that does not spoil the gauge invariance of the gravitational theory.  Finding 
such a description is an important next step in defining entanglement entropy for 
a gravitational theory.

There are a number of directions for future work on the extended phase space itself, outside 
of its application to entanglement entropy.  One topic of interest is to clarify the fiber bundle
geometry of the solution space $\SSS$, which arises due to diffeomorphism invariance.  
A fiber in this space consists of all solutions that are
related by diffeomorphism, and the $\dx^a$ fields define a flat connection
on the bundle.  Flatness in this case is equivalent to the equation $\del(\dx^a)+\frac12[\dx,\dx]^a
=0$ for the variation of $\dx^a$.  This fiber bundle description of $\SSS$ will be reported on
in a future work \cite{Jacobson2017}. 
Another technical question that arises is whether $\SSS$ truly carries a smooth manifold 
structure.  One obstruction to smoothness would be if the equations of motion are not well-posed
in some coordinate system.  In this case, the solutions do not depend smoothly on the initial
conditions on the Cauchy slice $\Sigma$, calling into question the smooth manifold structure
of $\SSS$.  If $X$ is used to define the coordinate system, this would mean that for
some values of $X$ the solution space is not smooth.  A possible way around this is to always work 
in a coordinate system in which the field equations are well-posed, and the gauge transformation
to this
coordinate system would impose dynamical equations on the $X$ fields.  Another obstruction 
to smoothness comes from issues related to ergodicity and chaos in totally constrained systems
\cite{Dittrich2015}.  It would be interesting to understand if these issues are problematic 
for the phase space construction 
given here, and whether the $X$ fields ameliorate any of these problems.

Another interesting application would be to formulate the first law of black hole mechanics 
and various related ideas in terms of the extended phase space.  This could be particularly 
interesting in clarifying certain gauge dependence that appears when looking at  
second order perturbative identities, such as described in \cite{Hollands2013}.  The edge modes
should characterize all possible gauge choices, and they may inform some of the relations 
found in \cite{Lashkari2016, Beach2016a, Faulkner2017} when considering 
different gauges besides the Gaussian null coordinates 
used in \cite{Hollands2013}.  They could also be useful in understanding
quasilocal gravitational energy, and in particular how to define the gravitational energy inside a 
small ball.  This can generally be determined by integrating a pseudotensor over the ball, but 
there is no preferred choice for a gravitational pseudotensor, so this procedure is ambiguous.  
It would be interesting if a preferred choice presented itself by considering second order variations
of the first law of causal diamonds \cite{Jacobson2015a, Bueno2017}, 
using the extended phase space.
Some ideas in this
direction are being considered in \cite{Jacobson2017b}, but it is difficult to find a quasilocal
gravitational energy that satisfies the desirable property of being proportional to the Bel-Robinson
energy density in the small ball limit \cite{Szabados2009, Senovilla2000a}.

Finally, it would be very useful to recast the extended phase space construction in 
vielbein variables.  
Some progress on the vielbein formulation was reported in \cite{Geiller2017}.  
Since vielbeins have an additional internal gauge symmetry associated with local 
Lorentz invariance,  care must be taken when applying  covariant
canonical constructions \cite{Jacobson2015b, Prabhu2017}.  It would be particularly interesting
to analyze the surface symmetry algebra that arises in this case, which could differ
from the algebra derived using metric variables because the gauge group is different.  
Comparing the algebras and edge modes in both cases would weigh on the question of 
how physically relevant
and universal their contribution to entanglement entropy is.

\section*{Acknowledgments}
I would like to thank William Donnelly, Laurent Freidel, Ted Jacobson, Eric Mintun and Arif Mohd
for helpful discussions, and Daniel Brennan for comments on a draft of this work.  
I thank the organizers of the TASI 2017 Summer School, where 
a portion of this work was completed. This research is supported by
the National Science Foundation under grant No. PHY-1407744.

\appendix
\section{List of identities} \label{app:ids}
This appendix gives a collection of identities for the exterior calculus on solution space $\SSS$
along with their proofs.  
\setenumerate[1]{label=\thesection.\arabic*}

\begin{enumerate}
\item $L_V = I_V \delta + \delta I_V$ \label{id:LV}
\begin{proof}
This follows from standard treatments of the exterior calculus \cite{Lang1985}.
\end{proof}

\item $L_V I_U = I_{[V, U]} + I_U L_V$ \label{id:LVIU}
\begin{proof}
This is simply the derivation property of the Lie derivative applied to all tensor fields on 
$\SSS$.   $I_U\alpha$ is a contraction of the vector $U$ with the one-form $\alpha$, so 
the Lie derivative first acts on $U$ to give the vector field commutator $L_V U = [V,U]$,
and then acts on $\alpha$, with the contraction $I_U$ now being applied to 
$L_V\alpha$.  Hence,
on an arbitrary form, $L_V I_U \alpha = I_{[V,U]}\alpha + I_U L_V\alpha$.
\end{proof}

\item $L_V Y^*\alpha = Y^*(L_V\alpha + \lie_{(I_V\dy)}\al)$ \label{id:LVY*a}
\begin{proof}
The discussion of section \ref{sec:cps} derived  equation
(\ref{eqn:LphY*a}), so all that remains is to show that $\dar^a(Y;V)$ is linear in the vector $V$.
This can be demonstrated inductively on the degree of $\alpha$.    For scalars, it is enough 
to show it holds on the functions $\phi^x$.  Applying \ref{id:LV}, we have on the one hand
\beq
L_V Y^*\phi = I_V \del Y^*\phi,
\eeq
while on the other hand,
\beq
L_V Y^*\phi = Y^*\left(L_V\phi + \lie_{\dar(Y;V)} \phi \right) = I_V Y^*\del\phi + Y^*\lie_{\dar(Y;V)}
\phi
\eeq
since $I_V$ commutes with $Y^*$.  Equating these expressions, we find
\beq \label{eqn:Y*liedar}
Y^* \lie_{\dar(Y;V)} \phi = I_V\left(\del Y^*\phi - Y^*\del\phi  \right).
\eeq
Since the right hand side of this expression is linear in $V$, $\chi(Y;V)$ must be as well.  

Now suppose \ref{id:LVY*a} holds for all forms of degree $n-1$, and take $\alpha$ to be 
degree $n$.  Then for an arbitrary vector $U$, $I_U Y^*\alpha$ is degree $n-1$, so 
\beq
L_V I_U Y^*\alpha = Y^*\left(L_V I_U\alpha + \lie_{(I_V\dar_Y)} I_U\alpha\right) = I_{[V,U]}Y^*\alpha
+ I_UY^*\left(L_V\al + \lie_{(I_V\dar_Y)}\alpha \right),
\eeq
where identity \ref{id:LVIU} was applied along with the fact that $I_U$ commutes with $\lie_\xi$. 
On the other hand,
\begin{align}
L_VI_U Y^*\al = I_{[V,U]} Y^*\alpha + I_U L_V Y^*\alpha = I_{[V,U]} Y^*\alpha + I_U
Y^*\left(L_V\al + I_{\bar\dar(Y;V)}\al \right).
\end{align}
Since $U$ was arbitrary, equating these expressions shows that 
$\bar\dar^a(Y;V)=I_V\chi_Y^a$, showing that the formula holds for 
forms of degree $n$.
\end{proof}

\item $I_V \lie_{\dar_Y} = \lie_{(I_V\dar_Y)} - \lie_{\dy} I_V$ \label{id:IVliedy}
\begin{proof}
This is essentially the antiderivation property applied to $\lie_{\dy}$.  The spacetime Lie 
derivative $\lie_\dy$ acting on a tensor can be written in terms of $\dar_Y^a$ and its derivatives 
contracted with the tensor, where all instances of $\dar_Y^a$ appear to the left.  It is straightforward
to see that when $I_V$ contracts with $\dar_Y^a$ in this expression, the terms will combine into
$\lie_{(I_V\dy)}$, and since $I_V$ does not change the spacetime
tensor structure of the object it contracts, the remaining terms will combine into $-\lie_\dy I_V$,
with the minus coming from the antiderivation property of $I_V$. 
\end{proof}

\item $\delta Y^* \alpha = Y^*(\delta\alpha +\lie_\dy\alpha)$ \label{id:dY*a}
\begin{proof}
This may also be demonstrated inductively on the degree of $\alpha$.  For scalars, we 
simply note that equation (\ref{eqn:Y*liedar}) is valid for arbitrary vectors $V$, and since 
$\dar^a(Y;V) = I_V \dar_Y^a$, we derive $\del Y^*\phi = Y^*(\del\phi +\lie_{\dy}\phi)$.
Assume now \ref{id:dY*a} holds for all $(n-1)$-forms, and take $\alpha$ an $n$-form and 
$V$ an arbitrary vector.  Then
\begin{align}
I_V \del Y^*\al &= L_V Y^*\alpha - \del I_V Y^*\al \nonumber \\
&= Y^*\left(L_V\al + \lie_{(I_V \dy)}\al - \del I_V \al - \lie_{\dy} I_V\al\right) \nonumber \\
&= I_V Y^*(\del\al + \lie_{\dy}\al )
\end{align}
The first equality applies \ref{id:LV}, the second uses \ref{id:LVY*a} and the fact that $I_V Y^*\al$
is an $(n-1)$-form, and the last equality follows from \ref{id:LV} and \ref{id:IVliedy}.  Since
$V$ is arbitrary, this completes the proof.
\end{proof}

\item $\frac12[\dy,\dy]^a = \dar_Y^b\nabla_b\dar_Y^a$ \label{id:dardar}
\begin{proof}
This is a consequence of the formula for the commutator of two vectors, $[\xi,\zeta] = 
\xi^b\nabla_b \zeta^a - \zeta^b\nabla_b \xi^a$, along with the fact that since $\dar^a$ is an $\SSS$
one-form, it anticommutes with itself.  Alternatively, the formula may be checked by 
contracting with arbitrary vectors $V$ and $U$.  Letting $I_V\dar_Y^a = - \xi^a$ and $I_U \dar_Y^a
=-\zeta^a$, we have
\beq
I_VI_U\frac12[\dy, \dy]^a = I_V[\dy,\zeta]^a = [\zeta,\xi]^a = 
\zeta^b\nabla_b\xi^a-\xi^b\nabla_b\zeta^a = I_V I_U \dar_Y^b\nabla_b\dar_Y^a.
\eeq
\end{proof}

\item $\lie_\dy \lie_\dy = \lie_{\frac12[\dy,\dy]}$ \label{id:liedarliedar}
\begin{proof}
For ordinary spacetime vectors $\xi^a$ and $\zeta^a$, the Lie derivative satisfies \cite{Edelen2005}
\beq
\lie_\xi \lie_\zeta = \lie_{[\xi,\zeta]} + \lie_\zeta \lie_\xi.
\eeq
Since $\dar_Y^a$ are anticommuting, this formula is modified to 
\beq
\lie_\dy \lie_\dy = \lie_{[\dy,\dy]} - \lie_\dy \lie_\dy,
\eeq
from which the identity follows.  Note that \ref{id:dardar} provides a formula for $[\dy,\dy]^a$.
\end{proof}

\item $\lie_\xi (Y^{-1})^* = (Y^{-1})^*\lie_{Y^*\xi}$ \label{id:liexiYi}
\begin{proof}
This identity is a standard property of the Lie derivative, see e.g.\ \cite{Kolar1993}.
\end{proof}

\item $\lie_\dx i_\dx = \frac12(i_{[\dx,\dx]} + d i_\dx i_\dx - i_\dx i_\dx d)$ \label{id:liedxidx}
\begin{proof}
The identity for ordinary spacetime vectors $\xi^a$ and $\zeta^b$ \cite{Edelen2005}
\beq
\lie_\xi i_\zeta = i_{[\xi,\zeta]} + i_\zeta \lie_\xi
\eeq
along with the fact that $\dx^a$ are anticommuting gives
\begin{align}
\lie_\dx i_\dx &= i_{[\dx,\dx]} - i_\dx \lie_\dx \nonumber \\
&= i_{[\dx,\dx]} - i_\dx di_\dx - i_\dx i_\dx d \nonumber\\
&= i_{[\dx,\dx]} - \lie_\dx i_\dx + di_\dx i_\dx - i_\dx i_\dx d, 
\end{align}
and moving $-\lie_\dx i_\dx$ to the left hand side proves the identity.
\end{proof}

\item $\lie_\dx \theta + \del i_\dx L + \lie_\dx i_\dx L = d\left(i_\dx\theta +\frac12 i_\dx i_\dx L\right)$
\label{id:liedxth}
\begin{proof}
The first term in this expression is $\lie_\dx\theta = d i_\dx \theta + i_\dx d\theta$, which gives 
one of the terms on the right hand side of the identity, along with $i_\dx d\theta$.  Next we have
\beq
\del i_\dx L = i_{\del \dx} L - i_\dx \del L = -\frac12 i_{[\dx,\dx]}L - i_\dx d\theta,
\eeq
where we applied equation (\ref{eqn:ddx}) for $\del \dx^a$, and used that $\del L = d\theta$
on shell.  The $-i_\dx d\theta$ term cancels against the similar term appearing in $\lie_\dx \theta$,
so that the remaining pieces are 
\beq
-\frac12 i_{[\dx,\dx]}L + \lie_\dx i_\dx L = \frac12 d i_\dx i_\dx L,
\eeq
which follows from identity \ref{id:liedxidx} and $dL=0$.  Hence, the terms on the left of the 
\ref{id:liedxth} combine into the exact form $d(i_\dx\theta +\frac12 i_\dx i_\dx L)$.
\end{proof}

\item $[V, \hat\xi] = (I_V\delta \xi^a)\,\hat{}$ \label{id:Vhxi}
\begin{proof}
Here we can use that on local $\SSS$-scalars, $L_{\hat\xi}\phi = \lie_\xi\phi$.  Then
\begin{align}
L_{[V,\hat\xi]}\phi = L_VL_{\hat\xi} \phi - L_{\hat\xi} L_V \phi  = L_V \lie_\xi\phi - \lie_\xi I_V\del\phi
=\lie_{(I_V\del \xi)\,\hat{}}\,\phi = L_{(I_V\del\xi)\,\hat{}} \,\phi,
\end{align}
hence, $[V,\hat\xi] = (I_V\del\xi^a)\,\hat{}$.
\end{proof}

\item $L_{\hat\xi} = \lie_\xi +I_{\del\xi\,\hat{}}$ \label{id:Lhxi}
\begin{proof}
This formula is meant to apply to local functionals of the fields defined at a single spacetime point.
Since $I_{\del\xi\,\hat{}}$ annihilates scalars, it clearly is true for that case.  Then assume the 
formula has been shown for all $(n-1)$-forms, and take $\alpha$ to be an $n$-form.  For an 
arbitrary vector $V$, since
$I_V\alpha$ is an $(n-1)$-form, we have 
\begin{align}
I_V L_{\hat\xi}\al &= L_{\hat\xi}I_V\al - I_{[\hat\xi,V]}\al = \lie_\xi I_V\al + I_{\del\xi\,\hat{}}I_V\al 
-I_{[\hat\xi,V]}\al \nonumber \\
& = I_V(\lie_\xi \al + I_{\del\xi\,\hat{}}\,\al) - I_{(I_V\del\xi)\,\hat{}}\,\al - I_{[\hat\xi,V]} \al,
\end{align}
and the last two terms in this expression cancel due to identity \ref{id:Vhxi}. Since $V$ was 
arbitrary, we conclude that the identity holds for all $n$ forms, and by induction for all 
$\SSS$ differential forms.
\end{proof}

\item $L_\hdx = I_\hdx \delta - \delta I_\hdx$ \label{id:Lhdxdef}
\begin{proof}
This is essentially a definition of what is meant by $L_\hdx$.  The left hand side is the graded
commutator of the derivation $I_\hdx$ and the antiderivation $\del$, which defines the 
the antiderivation $L_\hdx$ \cite{Kolar1993}.
\end{proof}

\item $[V,\hdx] = (\del I_V \dx^a)\,\hat{} - [I_V\dx,\dx]\,\hat{}$ \label{id:Vhdx}
\begin{proof}
This follows from the defining relation of the bracket \cite{Kolar1993},
\beq
L_V L_\hdx - L_\hdx L_V = L_{[V,\hdx]}.
\eeq
Applied to $\phi$ and defining $\nu^a = -I_V\dx^a$, this gives
\begin{align}
L_{[V,\hdx]}\phi &= (L_V L_{\hdx}-L_{\hdx} L_V)\phi \nonumber \\
&=I_V\del \lie_\dx \phi -\del\lie_\nu\phi - \lie_\dx I_V\del\phi \nonumber \\
&= I_V(\lie_{\del\dx} \phi - \lie_{\dx} \del \phi) - \lie_{\del\nu}\phi 
- \lie_{\nu}\del \phi+ I_V\lie_\dx \del\phi \nonumber \\
&= (\lie_{[\nu,\chi]} - \lie_{\del\nu})\phi \nonumber \\
&= (L_{[\nu,\chi]\,\hat{}} - L_{\del\nu\,\hat{}} )\phi,
\end{align}
To get to the third line, the expression (\ref{eqn:ddx}) for 
$\del\dx^a$ was used.  We then conclude $[V,\hdx] = [\nu,\dx]\,\hat{} - \del\nu\,\hat{}$, proving
the identity.
\end{proof}

\item $L_\hdx = \lie_\dx - I_{\del\dx\,\hat{}}$ \label{id:Ldx}
\begin{proof}
The formalism of graded commutators developed in \cite{Kolar1993} is a useful tool in 
proving this identity.  Given two graded derivations $D_1$ and $D_2$, their graded
commutator $D_1 D_2 - (-1)^{k_1 k_2} D_2 D_1$ is another 
graded derivation, where $k_i$ are the degrees of the 
respective derivations, i.e.\ the amount the derivation increases or decreases the degree of 
the form on which it acts.  Hence, since $I_V$ and $L_{\hdx}$ are  derivations of degrees $-1$
and
$1$, they satisfy
\beq \label{eqn:IVLhdx}
I_V L_{\hdx} + L_{\hdx} I_V = -L_{\hat{\nu}} + I_{[\hdx,V]},
\eeq
where $-\nu^a = I_V \dx^a$.  Similarly, we have 
\beq\label{eqn:IVhddx}
I_V I_{\del \dx\;\hat{}} + I_{\del\dx\;\hat{}} I_V = I_{(I_V\del\dx)\,\hat{}} = I_{[\nu,\dx]\,\hat{}},
\eeq
where equation (\ref{eqn:ddx}) was used in the last equality. 

We then prove the identity through induction on the degree of the form on which it acts.  
It is true for scalars because $I_{\del\dx} \phi=0$.  Then suppose it is true for all $(n-1)$-forms,
and take $\alpha$ to be an $n$-form.  For an arbitrary vector $V$ we have
\begin{align}
I_VL_\dx\al &= I_{[\hdx,V]}\al - L_{\hat\nu}\al - L_\dx I_V\al\nonumber \\
&= I_{\del\nu\,\hat{}\,} \al - I_{[\nu,\dx]\,\hat\,}\al - \lie_\nu\al - I_{\del\nu\,\hat{}\,}\al
 - \lie_\dx I_V\al + I_{\del \dx} I_V \al \nonumber \\
 &= I_V(\lie_\dx \al - I_{\del\dx\,\hat{}\,}\al).
\end{align}
The first line employs equation (\ref{eqn:IVLhdx}), the second line uses identities \ref{id:Vhdx}
and \ref{id:Lhxi} as well as the fact that $I_V\al$ is an $(n-1)$-form, and the third line
employs equation (\ref{eqn:IVhddx}). Since $V$ is arbitrary, we conclude the identity holds
for all $n$-forms, which completes the proof.
\end{proof}

\end{enumerate}

\section{Edge mode derivatives in the symplectic form} \label{app:edge}
In this appendix, we derive the result advertised in section 
\ref{sec:eps}, that the 
symplectic form (\ref{eqn:OmS}) does not depend on second or higher derivatives of 
$\dx^a$.  Derivatives of $\dx^a$
appear in $\Om$ through the terms $\del Q_{\dx} + \lie_{\dx}Q_\dx$.  The Lie derivative
term may be expressed
\begin{align}
\lie_{\dx}Q_{\dx} &= L_{\hdx} Q_{\dx} + I_{{\del(\dx)} \,\hat{}}\,Q_{\dx} \nonumber  \\
&= I_\hdx \del Q_\dx-\del I_\hdx Q_\dx-Q_{\del(\dx)}  \nonumber \\
&= I_\hdx \qo_\dx +I_\hdx Q_{\del(\dx)}  +\del Q_\dx - Q_{\del(\dx)} \nonumber \\
&= \qo_\dx + I_{\hdx}\qo_\dx +Q_{[\dx,\dx]}.
\end{align}
These steps invoke the identities \ref{id:Ldx}, \ref{id:Lhdxdef} and equations (\ref{eqn:ddx}) 
and (\ref{eqn:Ixidar}), as well as the defining relation
(\ref{eqn:dQdX}) for  
$\qo_\dx$. Adding $\del Q_\dx = \qo_\dx -\frac12 Q_{[\dx,\dx]}$ to this yields
\beq\label{eqn:dQdx}
\del Q_{\dx} +\lie_\dx Q_\dx = 2 \qo_\dx+I_{\dx} \qo_\dx + \frac12Q_{[\dx,\dx]}.
\eeq
From the derivation property of  $I_\hdx$ acting on $\SSS$-forms and the identity 
$I_\hdx \dx^a = -\dx^a$, it follows that $\qo_\dx+I_\dx \qo_\dx=  \qo[\phi;\lie_\dx\phi]_\dx$,
so that (\ref{eqn:dQdx}) can equivalently be expressed
\beq \label{eqn:dQlieQ2}
\del Q_\dx +\lie_\dx Q_\dx = \qo[\phi;\del\phi]_\dx  + \qo[\phi;\lie_\dx \phi]_\dx + \frac12 Q_{[\dx,\dx]}.
\eeq

This expression is now amenable to determining  how the derivatives of $\dx^a$ appear.
Both $\qo[\phi;\lie_\dx\phi]_\dx$ and $Q_{[\dx,\dx]}$ contain second derivatives.  The relevant 
term in $\qo[\phi;\lie_\dx\phi]_\dx$ 
comes from the variation of the Christoffel symbol in (\ref{eqn:qoxi}),
which gives
\begin{align}
&\,-\ep_{ab}E^{abcd}\left(\nabla_c \nabla_{(d}\dx_{e)}+\nabla_e\nabla_{(d}\dx_{c)}
-\nabla_d\nabla_{(c}
\dx_{e)}\right) \dx^e \nonumber \\
=&\,-\frac12\ep_{ab}E^{abcd}\left(  \nabla_{c}\nabla_{e}\dx_d - \nabla_d\nabla_e\dx_c\right)
\dx^e + \text{n.d.} \nonumber \\
=&\, -\ep_{ab}E\indices{^a^b^c_d} \left(\nabla_{(c}\nabla_{e)} \dx^d\right) \dx^e + \text{n.d.}, 
\label{eqn:epEdddX}
\end{align}
where ``$\text{n.d.}$" represents terms with no derivatives acting on $\dx^a$.
This derivation invokes the 
antisymmetry of $E^{abcd}$ on $c$ and $d$, and collects all terms involving antisymmetrized
derivatives of $\dx_d$ into the $\text{n.d.}$ piece, since these can be replaced by a Riemann
tensor contracted with an undifferentiated $\dx_d$. 

Second derivatives of $\dx^d$ also appear in $\frac12Q_{[\dx,\dx]}$ through the 
$E\indices{^a^b^c_d}$ term in the equation (\ref{eqn:Qxi}) for the Noether charge.  This term 
evaluates to 
\begin{align}
&\, -\frac12\ep_{ab}E\indices{^a^b^c_d}\nabla_c[\dx,\dx]^d \nonumber  \\
&=\,-\ep_{ab}E\indices{^a^b^c_d} \nabla_c(\dx^e\nabla_e\dx^d) \nonumber \\
&=\, -\ep_{ab}E\indices{^a^b^c_d}\left(\dx^e\nabla_{(c}\nabla_{e)}\dx^d + \nabla_c\dx^e\nabla_e
\dx^d  \right) +\text{n.d.},
\end{align}
which uses identity \ref{id:dardar}.  When added to (\ref{eqn:epEdddX}), the second derivative
terms cancel since $\dx^e$ is an $\SSS$ one form, so 
$(\nabla_{(c}\nabla_{e)}\dx^d)\dx^e = -\dx^e\nabla_{(c}\nabla_{e)}\dx^d$.  This 
shows that (\ref{eqn:dQlieQ2}) does not depend on second derivatives of $\dx^d$.

\bibliographystyle{JHEP}
\bibliography{cps}

\end{document}